\documentclass[twocolumn]{aastex631}
\usepackage{gensymb}
\usepackage{amsmath}

\newcommand{\rlow}{RACS-low}
\newcommand{\rmid}{RACS-mid}
\newcommand{\rhigh}{RACS-high}
\newcommand{\nsigma}[1]{\ensuremath{#1\sigma}}

\begin{document}

\title{Radio afterglows from tidal disruption events: An unbiased sample from ASKAP RACS}

\correspondingauthor{Akash Anumarlapudi}
\email{aakash@uwm.edu}

\author[0000-0002-8935-9882]{Akash Anumarlapudi}
\affiliation{Department of Physics, University of Wisconsin-Milwaukee, P.O. Box 413, Milwaukee, WI 53201, USA}

\author[0000-0003-0699-7019]{Dougal Dobie}
\affiliation{Centre for Astrophysics and Supercomputing, Swinburne University of Technology, Hawthorn, VIC 3122, Australia}
\affiliation{ARC Centre of Excellence for Gravitational Wave Discovery (OzGrav), Hawthorn, Victoria, Australia}

\author[0000-0001-6295-2881]{David L. Kaplan}
\affiliation{Department of Physics, University of Wisconsin-Milwaukee, P.O. Box 413, Milwaukee, WI 53201, USA}

\author[0000-0002-2686-438X]{Tara Murphy}
\affiliation{Sydney Institute for Astronomy, School of Physics, University of Sydney, NSW, 2006, Australia}
\affiliation{ARC Centre of Excellence for Gravitational Wave Discovery (OzGrav), Hawthorn, Victoria, Australia}

\author[0000-0002-5936-1156]{Assaf Horesh}
\affiliation{Racah Institute of Physics, The Hebrew University of Jerusalem, Jerusalem, 91904, Israel}

\author[0000-0002-9994-1593]{Emil Lenc}
\affiliation{CSIRO Space and Astronomy, PO Box 76, Epping, NSW, 1710, Australia}

\author[0000-0002-4405-3273]{Laura Driessen}
\affiliation{Sydney Institute for Astronomy, School of Physics, University of Sydney, NSW, 2006, Australia}

\author[0000-0002-3846-0315]{Stefan~W.~Duchesne}
\affiliation{CSIRO Space and Astronomy, PO Box 1130, Bentley, WA, 6102, Australia}

\author[0009-0008-6396-0849]{Hannah~Dykaar}
\affiliation{Dunlap Institute for Astronomy and Astrophysics, University of Toronto, 50 St. George St., Toronto, ON M5S 3H4, Canada}
\affiliation{David A. Dunlap Department of Astronomy and Astrophysics, University of Toronto, 50 St. George St., Toronto, ON M5S 3H4, Canada}

\author[0000-0002-3382-9558]{B.~M.~Gaensler}
\affiliation{Department of Astronomy and Astrophysics, University of California Santa Cruz, 1156 High Street, Santa Cruz, CA 95064, USA}
\affiliation{Dunlap Institute for Astronomy and Astrophysics, University of Toronto, 50 St. George St., Toronto, ON M5S 3H4, Canada}
\affiliation{David A. Dunlap Department of Astronomy and Astrophysics, University of Toronto, 50 St. George St., Toronto, ON M5S 3H4, Canada}

\author[0000-0002-2801-766X]{Timothy J. Galvin}
\affiliation{CSIRO Space and Astronomy, PO Box 1130, Bentley, WA, 6102, Australia}
\affiliation{International Centre for Radio Astronomy Research - Curtin University, 1 Turner Avenue, Bentley, WA 6102, Australia}

\author[0000-0002-4440-8046]{Joe Grundy}
\affiliation{CSIRO Space and Astronomy, PO Box 1130, Bentley, WA, 6102, Australia}
\affiliation{International Centre for Radio Astronomy Research - Curtin University, 1 Turner Avenue, Bentley, WA 6102, Australia}

\author[0000-0002-2155-6054]{George Heald}
\affiliation{CSIRO Space and Astronomy, PO Box 1130, Bentley, WA, 6102, Australia}

\author[0000-0001-7464-8801]{Aidan~W.~Hotan}
\affiliation{CSIRO Space and Astronomy, PO Box 1130, Bentley, WA, 6102, Australia}

\author[0000-0002-8314-9753]{Minh Huynh}
\affiliation{CSIRO Space and Astronomy, PO Box 1130, Bentley, WA, 6102, Australia}

\author[0000-0002-9415-3766]{James~K.~Leung}
\affiliation{Dunlap Institute for Astronomy and Astrophysics, University of Toronto, 50 St. George St., Toronto, ON M5S 3H4, Canada}
\affiliation{Racah Institute of Physics, The Hebrew University of Jerusalem, Jerusalem, 91904, Israel}

\author[0000-0002-2819-9977]{David McConnell}
\affiliation{CSIRO Space and Astronomy, PO Box 76, Epping, NSW, 1710, Australia}

\author[0000-0002-3005-9738]{Vanessa A. Moss}
\affiliation{CSIRO Space and Astronomy, PO Box 76, Epping, NSW, 1710, Australia}
\affiliation{Sydney Institute for Astronomy, School of Physics, University of Sydney, NSW, 2006, Australia}

\author[0000-0003-1575-5249]{Joshua Pritchard}
\affiliation{Sydney Institute for Astronomy, School of Physics, University of Sydney, NSW, 2006, Australia}
\affiliation{CSIRO Space and Astronomy, PO Box 76, Epping, NSW, 1710, Australia}
\affiliation{ARC Centre of Excellence for Gravitational Wave Discovery (OzGrav), Hawthorn, Victoria, Australia}

\author{Wasim Raja}
\affiliation{CSIRO Space and Astronomy, PO Box 76, Epping, NSW, 1710, Australia}

\author[0000-0002-7329-3209]{Kovi Rose}
\affiliation{Sydney Institute for Astronomy, School of Physics, University of Sydney, NSW, 2006, Australia}
\affiliation{CSIRO Space and Astronomy, PO Box 76, Epping, NSW, 1710, Australia}

\author[0000-0001-6682-916X]{Gregory Sivakoff}
\affiliation{Department of Physics, University of Alberta, CCIS 4-181, Edmonton AB T6G 2E1, Canada.}

\author[0000-0003-0203-1196]{Yuanming Wang}
\affiliation{Centre for Astrophysics and Supercomputing, Swinburne University of Technology, Hawthorn, VIC 3122, Australia}
\affiliation{ARC Centre of Excellence for Gravitational Wave Discovery (OzGrav), Hawthorn, Victoria, Australia}

\author[0000-0002-2066-9823]{Ziteng Wang}
\affiliation{International Centre for Radio Astronomy Research - Curtin University, 1 Turner Avenue, Bentley, WA 6102, Australia}

\author{Mark~H.~Wieringa}
\affiliation{CSIRO Space and Astronomy, PO Box 76, Epping, NSW, 1710, Australia}

\author[0000-0003-1160-2077]{Matthew~T.~Whiting}
\affiliation{CSIRO Space and Astronomy, PO Box 76, Epping, NSW, 1710, Australia}



\begin{abstract}
Late-time ($\sim$ year) radio follow-up of optically-discovered tidal disruption events (TDEs) is increasingly resulting in detections at radio wavelengths, and there is growing evidence for this late-time radio activity to be common to the broad class of sub-relativistic TDEs. Detailed studies of some of these TDEs at radio wavelengths are also challenging the existing models for radio emission. Using all-sky multi-epoch data from the Australian Square Kilometre Array Pathfinder (ASKAP), taken as a part of the Rapid ASKAP Continuum Survey (RACS), we searched for radio counterparts to a sample of optically-discovered TDEs. We detected late-time emission at RACS frequencies (742-1032\,MHz) in five TDEs, reporting the independent discovery of radio emission from TDE AT2019ahk and extending the time baseline out to almost 3000\,days for some events. 
Overall, we find that at least $22^{+15}_{-11}$\% of the population of optically-discovered TDEs has detectable radio emission in the RACS survey, while also noting that the true fraction can be higher given the limited cadence (2 epochs separated by $\sim 3\,$ years) of the survey. 
Finally, we project that the ongoing higher-cadence ($\sim 2$\,months) ASKAP Variable and Slow Transients (VAST) survey can detect $\sim 20$ TDEs in its operational span (4\,yrs), given the current rate from optical surveys. 
\end{abstract}

\keywords{Radio transient sources (2008) --- Tidal disruption (1696) --- Extragalactic radio sources (508) --- Radio continuum emission (1340) --- Radio sources (1358)}

\section{Introduction} \label{sec:intro}

The discovery of tidal disruption events \citep[TDEs;][]{rees1988} thus far was initially dominated by X-ray surveys \citep{Halpern2004} and then by optical/ultraviolet (O/UV) surveys in more recent times \citep{velzen2021,Hammerstein2023,Yao2023}. At O/UV wavelengths, emission from TDEs has a characteristic blue continuum with hydrogen and/or helium emission lines\footnote{This is true for sun-like stars. In general, the spectral signature depends on the composition of the disrupted star.} and can be accurately modeled as a black body with temperatures peaking near UV wavelengths \citep{gezaritdereview}. Radio emission from TDEs, expected from the interaction of nascent jets or outflows, was initially detected only in a handful of TDEs and this initial sample was dominated by TDEs that were discovered at higher energies. It was estimated by \cite{katetdereview} that not all TDEs result in radio detections, with only $\sim$ 20\% of them being radio bright. The distribution of radio luminosities from this initial crop of TDEs indicated a dichotomy at radio wavelengths where the luminosity differed by 2--3 orders of magnitude \citep{katetdereview}. The more luminous events resulted from relativistic jetted TDEs in which the radio luminosity exceeded $10^{40}$\,erg/s, while the less luminous events were from TDEs with sub-relativistic outflows where the isotropic radio luminosity were around $10^{38}$\,ergs/s \citep{Zauderer2011,Alexander2016,katetdereview}. 

Shock-accelerated relativistic electrons produce radio emission from TDEs via the synchrotron mechanism \citep{katetdereview}. This can be due to external shocks driven by jets/outflows or unbound stellar debris into the circumnuclear medium \citep[CNM;][]{Zauderer2011,Alexander2016,Krolik2016} or due to internal shocks within the jet \citep{pasham2018}. By modeling the spectral and temporal evolution of the emission, one can estimate the jet/outflow properties, particularly the velocity of the ejecta, their launch time relative to the optical flare, and the energy injected into the CNM \citep{Granot2002,Duran2013,Matsumoto2023}. Continuous monitoring of events in which early time ($\sim$ days to weeks after the optical flare) radio emission was detected, like Swift J1644+57 \citep{Zauderer2011} and ASASSN-14li \citep{Alexander2016}, demonstrated that the radio emission can be very long-lived, until $\sim$ years after the disruption. 

However, there are TDEs like ASASSN-15oi and AT2018hyz, in which early-time radio observations resulted in null detections, yet continued monitoring of these events until late time ($\sim$ months to years after the optical flare) resulted in radio detections \citep{Horesh2021,Cendes2022}. This can be either due to a delay in the ejection of the outflow \citep{Cendes2022} or due to the viewing effects of an off-axis observer looking at a relativistic jet \citep{Matsumoto2023,Sfaradi2024}. In addition, \citet{Horesh2021} found a radio re-brightening in ASASSN-15oi, $\sim 4$\,years after the initial optical discovery. \cite{Horesh2021} and \cite{Cendes2022} showed that the radio light curve in both these events showed a rise/decline that is steeper than any of the current predictions. More recently, studying late-time radio activity in TDEs using a sample of 23 TDEs, \cite{Cendes2023} showed that the launch of outflow can be delayed, by as much as $\sim 700$\,days, which raises the question of whether the phenomenon of delayed ejection is common in TDEs and whether the current models are adequate for describing the observed emission in TDEs like these.

While large samples of TDEs are coming from ongoing optical surveys \citep{vanvelzen2020,gezaritdereview,Yao2023}, the discovery space is expanding.  Recent studies like those of \cite{Velzen2016,velzen2021,Ning2021} and \cite{Masterson2024} have discovered TDEs at infrared (IR) wavelengths using dust echoes from TDEs. Using the first two epochs of the Very Large Array Sky Survey [VLASS;][]{vlass}, \cite{Somalwar2023} produced an independent sample of six radio TDEs that are optically bright. A few TDEs in this sample showed lower blackbody temperatures ($T_{bb}$) and luminosities ($L_{bb}$) compared to the optically discovered TDEs, indicating TDEs occurring in dust-obscured environments and adding to the sample of radio-first TDE discoveries \citep{Anderson2020,ravi2022}.
Such independent TDE discoveries from highly dust-obscured regions at radio/IR wavelengths can help constrain the true rate of TDEs and resolve the tension between the observed rate and the expected rate from theoretical predictions \citep{gezaritdereview,katetdereview,Yao2023}. Using the first three years of data from the Zwicky Transient Facility \citep[ZTF;][]{ztf}, \cite{Yao2023} estimated a volumetric rate of $3.1^{+0.6}_{-1.0}\times 10^{-7}\, \rm Mpc^{-3}\, \rm yr^{-1}$ TDEs ($L_{bb} > 10^{43}$\,erg/s). Comparing the rate of thermal TDEs to Swift J1644-like X-ray events \citep{katetdereview} and AT2020cmc-like optical events \citep{Andreoni2022}, the relative rate of jetted TDEs is estimated to be less than one percent of the thermal TDEs. This implies that the observed rate of thermal plus jetted TDEs is still lower than the current theoretical prediction by an order of magnitude  \citep{gezaritdereview}.

All-sky radio surveys can be an extremely useful resource in discovering radio afterglows serendipitously. However, multi-epoch data can be crucial to separate emission related to the TDE to emission from any active galactic nucleus (AGN) which may be present.  In particular, high cadence surveys like the Australian SKA Pathfinder Variable and Slow Transients survey \citep[ASKAP VAST;][]{vast,vastpilot}, can be very fruitful in getting a well-sampled light curve\footnote{VAST has a cadence of 2 weeks--2 months depending on the sky position.} for a larger sample of TDEs where dedicated follow-up of every individual event may not be possible/practical \citep[see e.g.,][for a serendepitous discovery of an off-axis TDE afterglow candidate]{jamesorphanafterglow}. Motivated by this, we used the data from the Rapid ASKAP Continuum Survey (RACS; \citealt{racs,racscatalogpaper}), a multi-epoch all-sky survey (see Table~\ref{tab:obs} for survey details) to search for   radio emission from TDEs discovered at higher energies (O/UV/X-ray). We then studied the prospects of finding radio TDEs in the VAST survey by projecting the rates estimated from the fraction of TDEs that are radio bright in the RACS survey. 

An alternate approach of discovering TDEs by modeling the radio light curve evolution using existing models \citep[e.g.,][]{Nakar2010} is used by \cite{Dykaar2024} to independently discover TDE candidates at radio wavelengths. Our approach is different from the untargeted and model-dependent search of \cite{Dykaar2024}, yet complementary since we find afterglows from TDEs like ASASSN-15oi, AT2018hyz etc, in which the observed radio emission can not be easily explained by the existing models. Unlike dedicated follow-up campaigns that extensively monitor a given sample of TDEs \citep{Cendes2023,Somalwar2023}, our approach is different, in that we study the prospects of discovering TDEs serendipitously in all-sky surveys, and hence our data are sparser. We focus instead on the nature of the TDEs we detect at lower observing frequencies, their rates, and the implications and expectations for the VAST survey.

Our article is structured as follows: in Section \S\ref{sec:observations}, we detail our observations, surveys used in this study, and our data reduction methods. In Section~\S\ref{sec:search} we discuss our sample selection technique. We present our detections in Section~\S\ref{sec:results} and describe the properties of the individual candidates in Sections~\S\ref{sec:15oi} through \S\ref{sec:agns}. Finally, we discuss the implications of our detections in Section~\S\ref{sec:discussion} and projections for future surveys like VAST in Section~\S\ref{sec:rates}, before concluding in Section~\S\ref{sec:conclusions}. 

Throughout this work, we use the \cite{Planck} model of cosmology, with $H_0 = 67.4\, \rm km\,Mpc^{-1}\,s^{-1}$.

\section{Observations and Data Analysis} \label{sec:observations}

\subsection{Rapid ASKAP Continuum Survey (RACS)}\label{sec:surveys}

The primary data set used in this work comes from all-sky  887.5\,MHz radio observations taken as a part of RACS --- \rlow. \rlow\ has been conducted at two separate epochs thus far, separated by $\sim$ 3\,years. In addition, RACS has also been conducted at two other frequencies, as single (so far) epoch surveys --- \rmid\ \citep[1367\,MHz; ][]{racsmid} and \rhigh\ {1655\,MHz; (\textit{in prep.})}, data from which we have used to study the behavior of the TDEs that we detected in \rlow. Details of each of these surveys are provided in Table~\ref{tab:obs}. 

Observations for all of the RACS surveys were carried out between March 2019 and April 2022. Data were processed using standard techniques recommended for ASKAP data \citep{askap}, using the ASKAP\textsc{soft} package \citep{askapsoft}, to generate both the images 
and the noise maps. A more detailed description of reduction techniques is provided by \cite{racs}. In this paper, we only used the total intensity (Stokes I) maps.  

\begin{deluxetable*}{lccccc}
\tablecaption{Survey details of all the different surveys used as a part of this article.\label{tab:obs}}
\tablehead{\colhead{Survey}  & \colhead{\rlow\ (epoch 1)}  & \colhead{\rlow\ (epoch 2)}  & \colhead{\rmid} & \colhead{\rhigh} & \colhead{VLASS}}
\startdata
Center frequency (MHz)  & 887.5  & 887.5 & 1367.5 & 1655.5 & 3000 \\
Bandwidth (MHz) & 288 & 288  & 144  & 200 & 2000 \\
Sky coverage & $-$90\degree $< \delta <$ +41\degree & $-$90\degree $< \delta <$ +51\degree & $-$90\degree $< \delta <$ +49\degree & $-$90\degree $< \delta <$ +48\degree & $-$40\degree $< \delta <$ +90\degree \\
Integration time & 15\,min & 15\,min & 15\,min & 15\,min & 5\,s\\
Median noise (mJy/beam)   & 0.25 & 0.19 & 0.20 & 0.19 & 0.12 \\
Angular resolution & $\sim$15\arcsec & $\sim$15\arcsec & $\sim$10\arcsec & $\sim$8\arcsec & 2.5\arcsec\\                     
Observations & $\sim$ March 2019 & $\sim$ March 2022 & $\sim$ January 2021 & $\sim$ December 2021 & \nodata\tablenotemark{a} \\
Instrument & ASKAP & ASKAP & ASKAP & ASKAP & VLA \\
Reference & \cite{racs} & in prep. & \cite{racsmid} & in prep. & \cite{vlass} \\
\enddata
\tablenotetext{a}{The first two epochs of VLASS were completed roughly in 2019 and 2021, and the third observing run is currently ongoing.}
\end{deluxetable*}

\subsection{Variable and Slow Transients Survey (VAST)}

VAST \citep{vast,vastpilot} is a radio survey that will image almost one-quarter of the entire sky repeatedly for 4\,years. VAST is divided between the Galactic and extra-galactic sky with the Galactic sky being observed with a cadence of roughly 2\,weeks and the extra-galactic sky with a cadence of roughly 2\,months. VAST pilot surveys \citep{vastpilot} were carried out in-between the two \rlow\ epochs, and the main VAST survey\footnote{\url{https://www.vast-survey.org/Survey/}} began its operation in December 2022. For the TDEs that we detected in the \rlow\ dataset, we augmented the RACS data with data from the VAST survey, if the transient falls inside the VAST footprint. The survey parameters of VAST are similar to the \rlow\ (see Table~\ref{tab:obs}) survey, except for a 12\,min integration time per field in VAST compared to a 15\,min observation in RACS. A more detailed description of the pilot and the full surveys is provided by \cite{vast,vastpilot,jamesorphanafterglow,joshuaradiostars}.  

\subsection{VLA Sky Survey (VLASS)}
In addition to the RACS and VAST survey data, we also made use of the VLA Sky Survey \citep[VLASS;][]{vlass}. VLASS is an all-sky survey\footnote{North of $-40\degree$ declination.} spanning 2-4\,GHz and plans to scan the entire sky at three different epochs with a cadence of roughly 32\,months between the epochs. The first two epochs have been completed and the third epoch is underway. For the TDEs detected in \rlow\ data, we used the VLASS quick-look images\footnote{\url{https://archive-new.nrao.edu/vlass/quicklook/}} \citep{vlass} to measure the flux density at 3\,GHz. 

\subsection{Search methodology}\label{sec:search}
We selected all the TDEs from the transient network server (TNS)\footnote{\url{https://www.wis-tns.org/}} that were spectroscopically classified as TDEs, as well as those that were optically discovered in all-sky surveys like the ZTF and All-Sky Automated Survey for Supernovae \citep[ASAS-SN\footnote{\url{https://www.astronomy.ohio-state.edu/asassn/}};][]{asassn}, which resulted in 63 events \citep{Auchettl2017,Hammerstein2023,Yao2023}. We then discarded 13 events that are outside the \rlow\ epoch 1 footprint, as well as those events where the optical discovery occurred after \rlow\ epoch 2, leaving 43 events in our sample. We examined the \rlow\ total intensity (Stokes~I) sky maps to look for radio emission at the TDE positions. Radio emission in TDEs can be observable $\sim$years after the initial disruption\footnote{The radio emission can persist for $\sim$ years after the optical flare \citep[like Swift J1644;][]{Zauderer2011} in a few TDEs, but is only observable at late times \citep[like ASASSN-15oi;][]{Horesh2021} in a few others.} \citep[see][]{katetdereview,Cendes2022,Sfaradi2024,Cendes2023} and hence we restricted our cross-match to spatial coincidence, relaxing any constraint on the temporal coincidence as long as the TDE was discovered before the second \rlow\ epoch. The positional accuracy for the ASKAP data is 2.5\arcsec\footnote{This is including the systematic component of the offset \citep[see][]{racs}.} and hence we used twice this as our search radius, 5\arcsec, when astrometrically crossmatching the TDEs. This resulted in 11 TDEs for which we detected coincident radio emission in \rlow. However, only 5 of these events showed significant variability in their light curve between the two \rlow\ epochs. The remaining 6 events did not show any significant evolution between the \rlow\ epochs, which made it difficult to rule out underlying host galaxy/host AGN emission (see Section~\S\ref{sec:agns} for more details). 


In the five TDEs with coincident variable radio emission, the emission lasted $\sim$\,years after the initial optical outburst, with the longest radio lived TDE lasting $\sim 8$\,yrs. Our detections add to the sample of TDEs reported by \cite{Cendes2023}, where late-time radio emission is seen. However, only one TDE (AT2018hyz) is common between our sample and \cite{Cendes2023}. Table~\ref{tab:flux} gives the flux density measurements for all these events.
For all the TDEs that are in the RACS footprint, but resulted in non-detections we provide upper limits (3-$\sigma$) on the radio flux density and radio luminosity in Table~\ref{tab:limits}.

\section{Individual Tidal Disruption Events}\label{sec:results}

Given the nature of this study, our light curves are sparser than dedicated campaigns like those of \citet{Goodwin2022} or \citet{Cendes2022,Cendes2023}. We therefore make simplifying assumptions about the spectral and temporal properties of the observed emission to estimate the source properties. We modeled the late-time radio spectrum as a broken power-law with the break frequency corresponding to synchrotron self-absorption (SSA) frequency ($\nu_{\rm ssa}$), but adapted from \cite{Granot2002} to join the power-laws smoothly \citep[see case \textbf{2} of Figure \textbf{1} of][]{Granot2002}. We modeled the temporal evolution of the light curve using \cite{Chevalier1998}: a rising power-law when the emission is optically thick smoothly joined by a declining power-law when the emission becomes optically thin.

To infer source parameters we assume that the energy stored in magnetic fields is similar to the energy of the relativistic electrons, (\textit{equipartition}; \citealt{Pacholczyk1970}). Since the time scale of our radio detections is $\approx$ year(s), unless we see evidence for on-axis jets (radio luminosity consistent with Swift J1644 or AT2020cmc-like events) or off-axis relativistic jets (characterized by steep rise time), we assume that the bulk Lorentz factor is close to 1 (Newtonian case). We assume that roughly 10\% of the energy in heavy particles is used to accelerate the electrons to relativistic speeds ($\epsilon_e \approx 0.1$). Assuming a power-law seed electron energy distribution $N(E) dE = A E^{-p}$, with $p$ being the index, we infer the emission radius ($R_{\rm eq}$) and the total equipartition energy following \cite{Duran2013}\footnote{\textbf{Since the peak frequency corresponds to $\nu_{ssa}$, we correct the total equipartition energy by accounting for the radiation emitted at $\nu_m$.}}. We caution that the outflow geometry of sub/non-relativistic outflows can be quasi-spherical or asymmetrical, in which the filling factors can differ, but as noted by \cite{Pacholczyk1970}, the estimated source properties are relatively insensitive to these. Hence, in this work, we assume that the geometry is nearly spherical. Further, we assume that the observed radio emission arises from a thin shell of expanding outflow (of width $\approx 0.1$R, where R is the radius, e.g., \citealt{Alexander2016}) and is spherically symmetric. For such cases, the areal and volume filling factors, $f_A$, and $f_V$ \citep[see][]{Duran2013} are 1 and 0.36 respectively.

\subsection{ASASSN-15oi}\label{sec:15oi}
\begin{figure}
    \centering
    \includegraphics[scale=0.4]{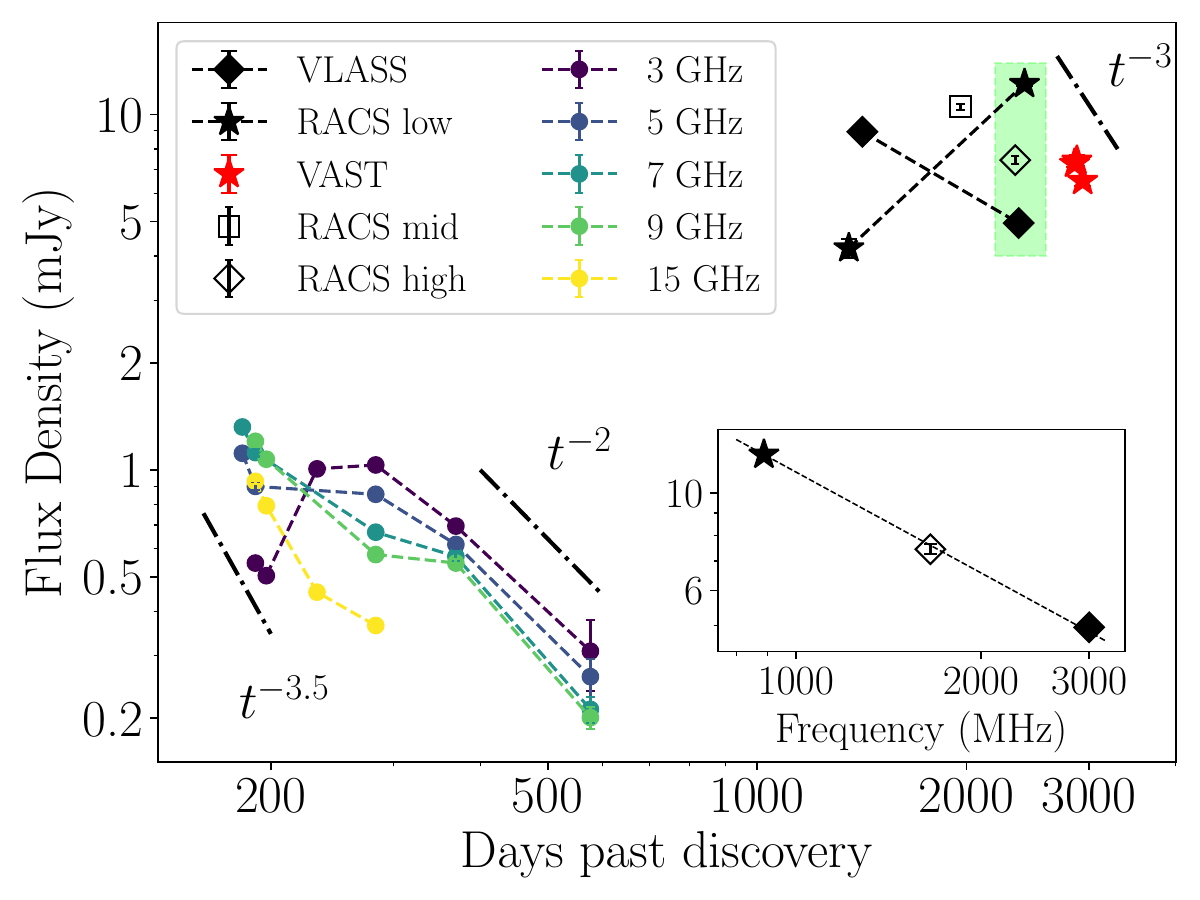}
    \caption{Lightcurve of the TDE ASASSN-15oi using RACS, VLASS, and archival data. In all the figures in this article, \rlow\ data is shown as black stars, \rmid\ as open black squares, \rhigh\ as open black diamonds, and VLASS data as filled black diamonds. RACS data combined with the data from the VAST full survey, shown as red stars, reveal the rise and decline of the 888\,MHz light curve. Shown as multicolored dots is the archival lightcurve, from 3\,GHz to 15\,GHz, adapted from \cite{Horesh2021}. The green stripe shows the data used to estimate the electron distribution index. The inset plot shows this spectrum and the dashed line in the inset plot shows the best-fit power law to these data (see text for more details). Shown as dashed-dotted lines are the visual guides for different power-law declines.}
    \label{fig:15oi}
\end{figure}

After an initial non-detection at radio wavelengths (up to $\sim$ 6\,months), \cite{Horesh2021} reported the discovery of a radio counterpart to ASASSN-15oi \citep{Holoien2016} that rose steeply ($\sim t^{4}$). This was followed by a steep fall (steeper than $t^{-3}$) that became shallower at late times (see Figure~\ref{fig:15oi}). \cite{Horesh2021} noted that such steep rise and fall times could not be explained by a standard forward shock and CNM interaction model. \cite{Horesh2021} also reported a very late time re-brightening ($\sim$1000\,days later) in the VLASS epoch 1 data. 

We detected very late time re-brightening in \rlow\ data and the lightcurve continued to rise (roughly as $\sim t^2$) and peaked $\sim 2500$~days after the optical flare (see Figure~\ref{fig:15oi}). This very late time re-brightening was replicated in the \rmid\ and \rhigh\ data as well. VAST observations for this transient revealed that the emission started to decline steeply ($\sim t^{-3}$) following the peak. This very late time decline is similar to the behavior that \cite{Horesh2021} reported following the initial radio peak. As \cite{Horesh2021} points out, the changes in various decline rates of emission could point to changes in the CNM density profile or a structured jet. However, it is difficult to reconcile such steep rise and fall times with the existing afterglow models. 

Using the second epoch of VLASS observations, we find that the 3\,GHz light curve is declining, roughly following a $t^{-1}$ decline (see Figure~\ref{fig:15oi}). This is in contrast to the rising 887.5\,MHz light curve during the same period, which suggests that the emission at 3\,GHz was optically thin during this period and that at 887.5\,MHz was optically thick. This can be explained by the peak frequency gradually transitioning to lower frequencies at late times, a trend that is expected and was also observed by \cite{Horesh2021} during the initial radio observations. This is also consistent with our 887.5\,MHz observations, which revealed a turnover indicative of emission transitioning from optically thick to thin at $>3000$ days.

The \rhigh, VLASS epoch 2, and the \rlow\ epoch 2 data are separated by $\sim 75$\,days, and under the assumption that the spectral evolution during this time frame is minimal (given the active cycle of $>4$\,years), we found that the spectrum at this epoch ($\sim 2400$\,d after the event) is well fit by a power law (with the spectra index, $\alpha=-0.75\pm 0.2$, where $S_\nu \propto \nu^\alpha$). We assumed that the self-absorption frequency is closer to the RACS observing frequency (887.5\,MHz)\footnote{The radio emission being optically thick at 887.5\,MHz at this time (which continues until $\sim 2800$\,days after the disruption) and thin at 3\,GHz partially supports this assumption.}, without attempting a physical model for the origin of this\footnote{\cite{Horesh2021} found that the initial radio spectrum showed large deviation from the SSA spectrum in the self-absorbed part, but might be consistent with free-free absorption.}, and estimate the electron distribution index $p=2.5\pm 0.2$. Given the peak frequency $\nu_{\rm p} \approx 887.5$\,MHz, the peak flux density $F_{\nu, \rm p}=12.2$\,mJy, and $p=2.5$, we derive a lower limit of $R_{\rm eq} \approx 6\times 10^{17}$\,cm, on the emission radius and $E_{\rm eq} \approx 1\times 10^{50}$\,erg on the total energy.  

\subsection{AT2019ahk / ASASSN-19bt}\label{sec:19ahk}

\begin{figure}
    \centering
    \plotone{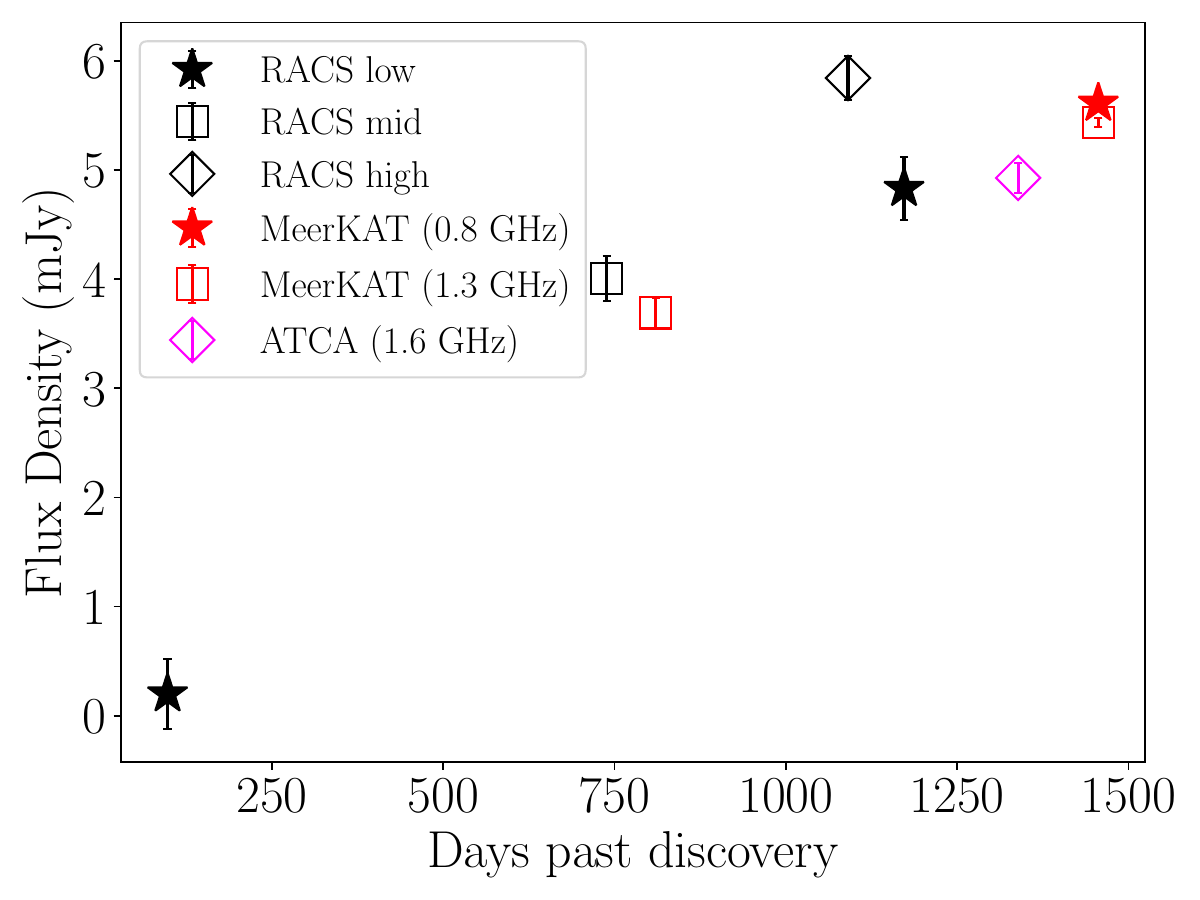}
    \caption{Lightcurve of the TDE AT 2019ahk confirming the late-time rise of radio light curve at all the RACS observing frequencies. Also shown are the contemporary data at 1.3\,GHz and 1.6\,GHz adapted from \cite{Christy2024} showing the continued rise of 0.8-0.9\,GHz light curve, turn over of 1.6\,GHz light curve. This source is too far south for VLASS.}
    \label{fig:19ahk}
\end{figure}

AT2019ahk was discovered as an optical transient by \cite{Holoien2019}. We report an independent radio discovery of this event
in RACS data at all three frequencies (see Figure~\ref{fig:19ahk}), where we saw a rising transient over 3\,years. \cite{Christy2024} reports archival radio detection of AT2019ahk roughly 4\,years before the disruption and estimates underlying host galaxy emission to follow $F_{\nu, \rm host} = 0.439\, (\nu/2.1)^{-1}\, \mathrm{mJy}$. Combining the RACS data with contemporaneous data from \cite{Christy2024}, we see that the 0.8--0.9\,GHz light curve is still rising $\approx 1500$ days after the event, but the 1.6\,GHz light curve started to decline. This hints that emission at 1.6\,GHz has transitioned to an optically thin regime, but emission at lower frequencies is still optically thick. Hence, the SSA frequency is very close to \rhigh\, the frequency at $\approx 1100$ days, consistent with the peak frequency estimated by \cite{Christy2024}. Using $p \approx$ 2.7 (using existing literature, e.g., \citealt{Goodwin2022,Cendes2022,Cendes2023} and also consistent with \citealt{Christy2024}), we estimate the equipartition emission radius for $\nu_{\rm p}=1.655$\,GHz, $F_{\nu, \rm p}=6.4$\,mJy to be $\approx 1 \times 10^{17}$\,cm and total energy to be $\approx 7 \times 10^{48}$\,erg at $\delta t=1100$\,days. 

\subsection{AT2019azh}\label{sec:19azh}
\begin{figure}
    \centering
    \plotone{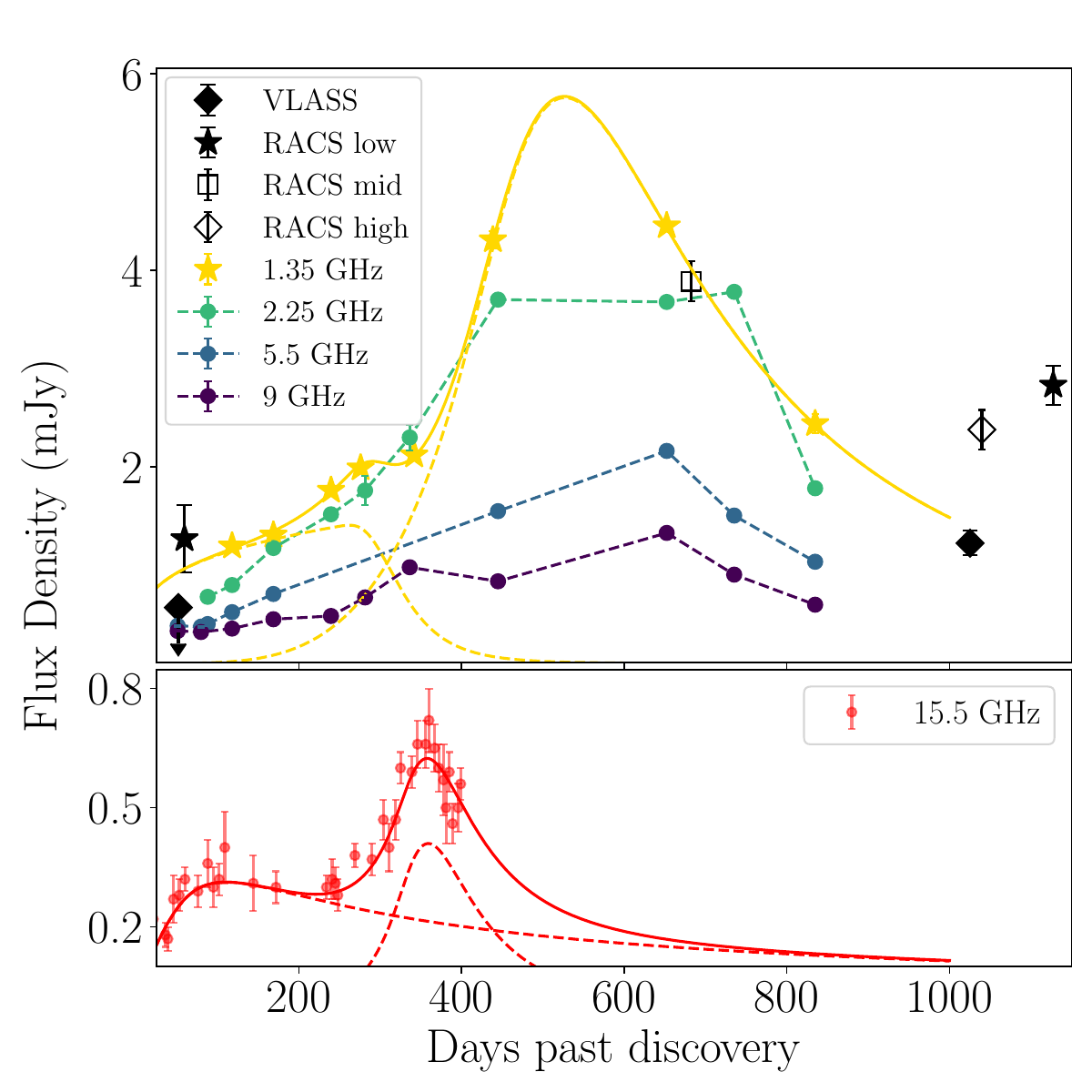}
    \caption{\textit{Top panel:} The light curve using archival data \citep[from][]{Sfaradi2022, Goodwin2022} is shown at 4 different frequencies (1.3, 2.25, 5.5, 9\,GHz). The yellow curve is the best-fit model for the 1.4\,GHz data (see text for the model), with the dashed lines representing the individual components and the solid line indicating their sum. \textit{Bottom panel:} 15.5\,GHz lightcurve for reference using the data from \cite{Sfaradi2022}. The red curve is the two-component model proposed by \cite{Sfaradi2022}, with the dashed line representing the individual components and the solid line, their sum.}
    \label{fig:2019azh}
\end{figure}

Using multi-frequency observations of multiple epochs, \cite{Goodwin2022} modeled the radio spectrum of AT2019azh to find a free expansion of the ejecta that showed signs of deceleration post $\sim 450$\,days of the disruption. \cite{Sfaradi2022}, on the other hand, modeled the 15.5\,GHz lightcurve and found evidence for two emission components (see Figure~\ref{fig:2019azh}), which led the authors to propose a state transition similar to the ones observed in X-ray binaries. 

We found this TDE in the \rlow\ data as a slowly rising source, increasing by a factor of $\sim 2$ between the two epochs. We also detected this source in the \rmid\ and \rhigh\ data sets. Using the \rmid\ data and the data from \cite{Goodwin2022}, we modeled the 1.4\,GHz lightcurve reasonably well by a two-component model similar to \cite{Sfaradi2022}. Figure~\ref{fig:2019azh} shows the full light curve for this event where the similarity can be seen between the shapes of the 15.5\,GHz and 1.4\,GHz light curves, although the rise and fall times at these frequencies are different. At 1.4\,GHz, the two components rose to a peak at $\sim 300$ and 520\,days respectively, slower than the 15.5\,GHz light curve that took 130\,days and 360 to rise. This is broadly consistent with the underlying model \cite{Chevalier1998} where the emission at different frequencies is self-similar but the emission at lower frequencies has longer rise times.

However, the very late time ($\gtrsim 3$\,yrs) relative behavior between the \rmid\ and the \rhigh\ data is puzzling. \cite{Goodwin2022} noted that the peak frequency at late times was $<1$\,GHz, which meant that the spectrum above this should be a declining one. But the observed flux density in \rhigh\ is higher than the model-predicted flux density in \rmid\ by a factor that is roughly consistent with the SSA mechanism (where $S_\nu\propto \nu^{5/2}$). This might be indicative of the peak frequency increasing to higher frequencies at late times, something that \cite{Cendes2022} observed in another event, AT2018hyz, indicative of late-time source activity. The \rlow\, second epoch detection postdates this, however the lack of continued coverage through very late times makes it difficult to distinguish whether this consistent with the initial decline, or is a signature of very late time rebrightening, as hinted by the \rhigh\ data.

The \rhigh\ and VLASS observations are separated by $\approx 14$\,days; under the assumptions that i) this interval is much shorter than the evolutionary timescale of the radio emission, and ii) the peak frequency rose, but to a value lower than the observing frequency of \rhigh. We estimated the electron distribution index to be $p=3.2\pm 0.4$ (at $\Delta t=1030$\,days), consistent with the electron distribution of \cite{Goodwin2022} at late times (849\,days). This seems to hint that the emission we see at very late times might still be coming from the same family of electrons. 

\subsection{AT2018hyz}\label{sec:18hyz}

\begin{figure}
    \centering
    \includegraphics[scale=0.4]{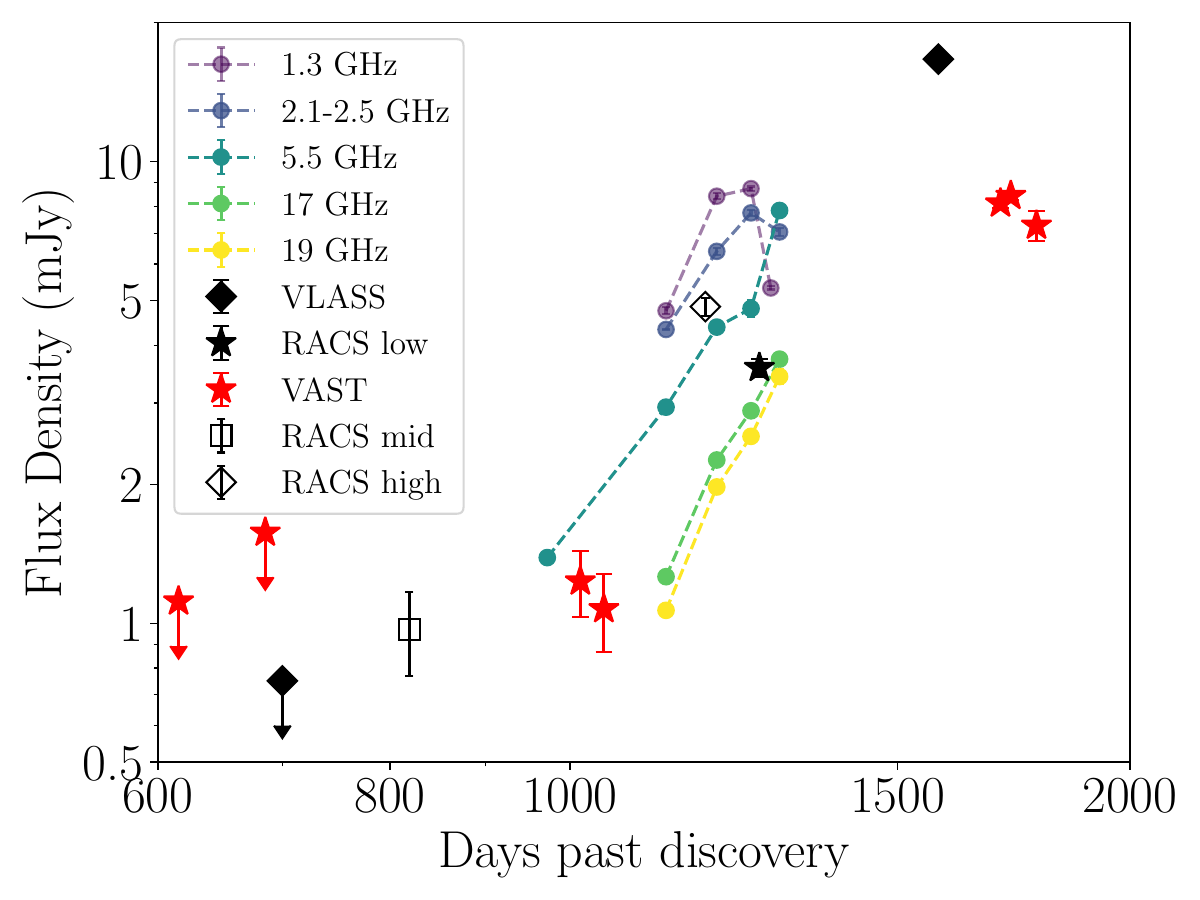}
    \caption{Light curve of TDE AT2018hyz using RACS/VAST data, VLASS data, and archival data \citep[from][]{Cendes2022}. The archival lightcurve, at 5 different frequencies from 1.3\,GHz to 19\,GHz, adapted from \cite{Cendes2022}, is shown as multi-colored dots for reference.}
    \label{fig:2018hyz}
\end{figure}

AT2018hyz was first detected at radio wavelengths $\sim2.5$\,years after the optical outburst \citep{Horesh2018hyzATel}, and showed an unusually steep rise ($\sim t^{4-6}$) at most of the observed frequencies (1.3-19\,GHz) \citep{Cendes2022,Sfaradi2024}. \cite{Cendes2022} noted that the light curve at lower frequencies ($\lesssim 3$\,GHz) began to decline (see Figure~\ref{fig:2018hyz}) at the end of their observing campaign ($\sim 1250$\,d past optical outburst). Modeling the spectrum at multiple epochs, \cite{Cendes2022} also found that the peak frequency increased roughly from 1.5\,GHz to 3\,GHz at late times. However, following the off-axis jet model proposed by \cite{Matsumoto2023}, \cite{Sfaradi2024} showed that the observed radio emission in AT2018hyz is also consistent with late-time emission from a narrow jet ($\sim 7\degree$) as viewed by an off-axis observer ($\sim 42\degree$). 

Upon finding this source in \rlow\ data, we looked at the detailed VAST light curve and found no discernible radio emission until late times and a very steep rise at late times ($\sim t^4$ rise; see Figure~\ref{fig:2018hyz}), both of which were consistent with \cite{Cendes2022} and \cite{Sfaradi2024}. We also found that the  887.5\,MHz emission continued to rise until our final observation ($\Delta t=1700$\,days)\footnote{Using the data from our latest observation at 1757 days, we find a hint of a turn-over in the 887.5\,MHz lightcurve, but we need additional data to robustly confirm this.}. However, given the steep rise of this particular transient and the gap between \rlow\ epoch 2 and the VAST full survey data, we cannot rule out a decline seen by \cite{Cendes2022} at frequencies below $3$\,GHz, followed by a rebrightening at 887.5\,MHz instead of a single brightening episode.

We then investigated the sudden jump in the peak frequency from 1.5\,GHz to 3\,GHz reported by \cite{Cendes2022} \citep[see Figure 3 and Section 4.1 of][]{Cendes2022}. At day 1251, \cite{Cendes2022} found that the peak frequency is 1.5\,GHz but the data used in this fit were all at frequencies $>$1.12\,GHz, where the self-absorbed part of the spectrum might not have been well captured. Combining the 887.5\,MHz RACS data from day 1263 with the data from day 1251, we found that the peak frequency rose to 1.9\,GHz, as opposed to 1.5\,GHz. At this epoch, we also find that the absorption part of the spectrum is more or less consistent ($S_\nu \sim \nu^{2.7}$) with what is expected from the SSA mechanism ($S_\nu \propto\nu^{5/2}$). This rise in the peak frequency to roughly 3\,GHz at day 1282 might be explained by this gradual increase in the peak frequency rather than a sudden shift, something similar to what we found in AT2019azh (see \S\ref{sec:19azh}). 

Using the latest epoch of VLASS data, we found that the 3\,GHz emission also rose from an early non-detection as $t^4$ (see Figure~\ref{fig:2018hyz}), consistent with the RACS/VAST data, to a remarkably bright 16.5\,mJy. This is consistent with the very late-time brightening of this transient in radio.

\subsection{AT2019qiz}\label{sec:19qiz}

\begin{figure}
    \centering
    \plotone{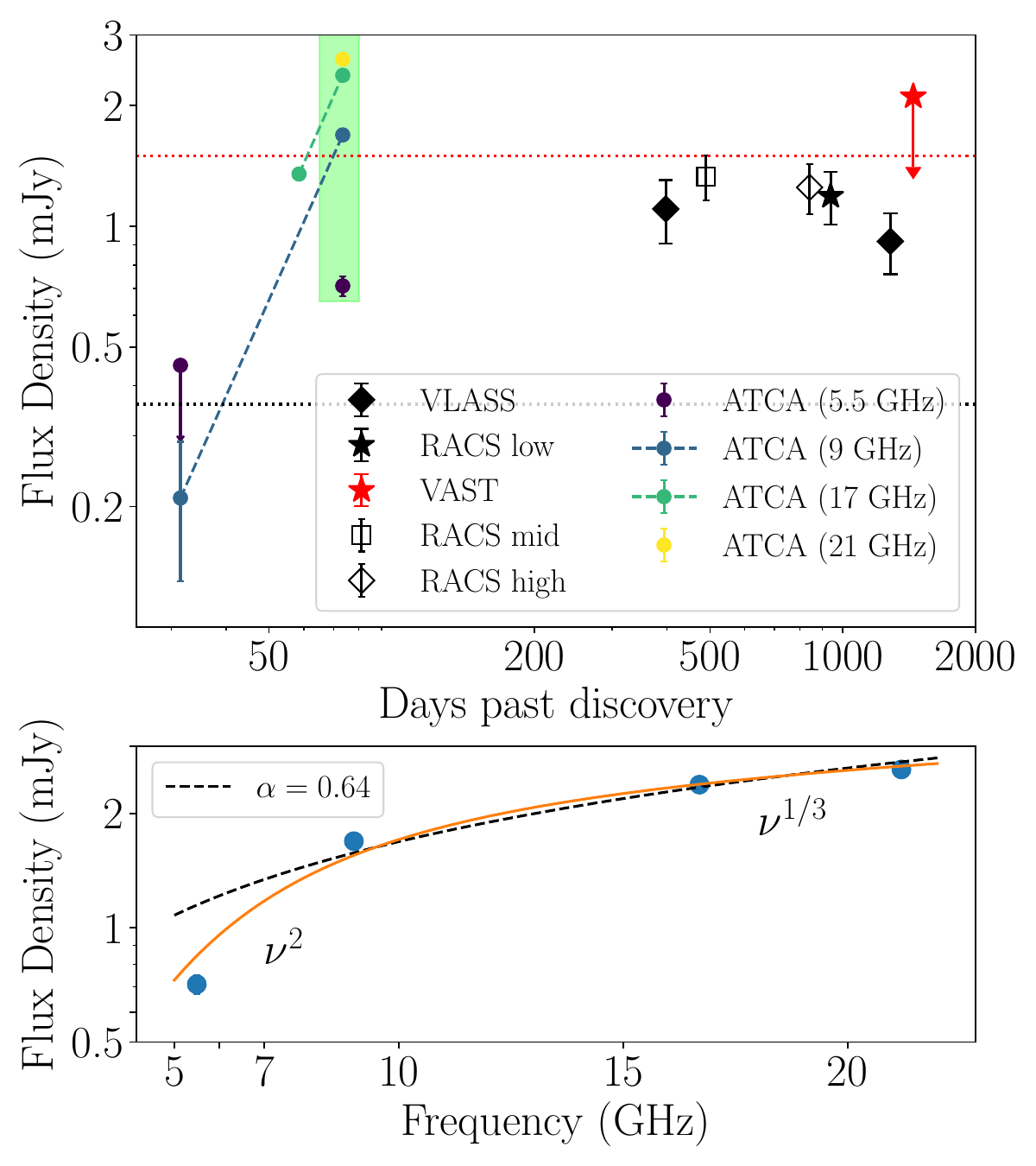}
    \caption{\textit{Top panel:} Light curve of the TDE AT2019qiz using RACS, VLASS, and archival data at 4 different frequencies from 5.5\,GHz to 21\,GHz from \cite{Andrew2019, Andrew20192}. The green stripe shows the multi-frequency epoch at $\sim 75\,$d which we use for spectral fitting. The black dotted line shows the 3-$\sigma$ pre-disruption limit from VLASS data and the red dotted line shows the same from \rlow\, epoch 1 data. \textit{Bottom panel:} Spectrum of this transient using observations at 4 different frequencies taken at the same epoch ($\sim 75\,$d post optical discovery; indicated by the box in the top panel). The orange line is the best fit broken power-law to the observed spectrum (see section \S\ref{sec:19qiz}). The dashed black line is the best fit single power-law spectrum (with spectral index $\alpha=$0.64).}
    \label{fig:2019qiz}
\end{figure}

AT2019qiz \citep{Siebert2019,Nicholl2020,Hung2021,Patra2022} has received comparatively very little follow-up at radio wavelengths, with \cite{Andrew2019,Andrew20192} presenting the initial radio detections that indicated a rising transient at multiple frequencies but with no robust analysis presented (see Figure~\ref{fig:2019qiz}).

We found this transient in \rlow\ data brightening from a non-detection in epoch 1 to a flux density level of $\sim 1$\,mJy  in the second epoch of \rlow, consistent with \rmid\ and \rhigh. This  suggests that this source might be very slowly evolving or that it may be steadily emitting at higher flux density levels. The VAST full survey data resulted in a non-detection, which indicated that the flux density variation was $<$30\% of the mean (see Figure~\ref{fig:2019qiz}). We also inspected the VLASS epoch 1 image that predated the optical disruption time and did not find a detection putting a \nsigma{3} upper limit of 0.36\,mJy on the persistent emission at 3\,GHz. However, the transient rose to persistent levels of 1\,mJy in the latter VLASS epochs (see Figure~\ref{fig:2019qiz}). 

Motivated by this behavior, we wanted to see if the early-time behavior was consistent with an afterglow or if it was different, in which case it might provide clues to the nature of the underlying emission. Using the data from \cite{Andrew2019,Andrew20192}, we found that the initial rise time estimated at different frequencies seems to be consistent with $t^{2.5}$ at both 17 and 9\,GHz. This $t^{2.5}$ increase was also consistent with the non-detection of this transient at 5.5\,GHz at early times. The spectrum at $\sim 75\,$d seemed to be inconsistent with a synchrotron self-absorption spectrum ($S_\nu \propto \nu^{5/2}$), so, we tried to model the break frequency as minimal frequency ($\nu_{m}$) instead of self-absorption frequency \citep[see spectrum \textbf{1} in Figure \textbf{1} of][]{Granot2002}. Here the two rising power-laws spectral indices are +2 (Rayleigh-Jeans tail) at frequencies below the break and +1/3 at frequencies above the break frequency. We found a reasonable fit to the spectrum in this case (see bottom panel of Figure~\ref{fig:2019qiz}) with the break frequency around $\sim 9$\,GHz. This is indicative that at early times the emission seems to be consistent with an afterglow.

We tried to reconcile with the late-time radio observations from RACS and VLASS. The lack of late-time evolution likely ruled out the scenario where the late-time activity was still dominated by the emission powered by the CNM interaction. It is also possible that there was prior nuclear activity (possibly from an AGN) in this galaxy which is visible once the transient faded away. The non-detection in VLASS and RACS data before the optical discovery makes this unlikely\footnote{In particular, the AGN flux density variation with respect to the VLASS non-detection in epoch 1  has to be at least a factor of $\sim 5-6$.}, but cannot be ruled out entirely. Although not temporally simultaneous, if we assume that the source is persistent and non-variable, the spectrum might be consistent with a flat spectrum at late times, using the RACS and VLASS data. It might be possible that a jet was launched at early times and we are looking directly into the emission from the jet at late times, which could explain the flat spectrum. If this were the case, then it might be an interesting situation in which late-time emission from the jet was directly seen and would add to the small sample of jetted radio TDEs, but given the sparsity of the data, it cannot be firmly established. 


\subsection{Steady Radio Sources: Probable AGN/Host Galaxy Emission}\label{sec:agns}
\begin{figure*}
    \centering
    \includegraphics[scale=0.5]{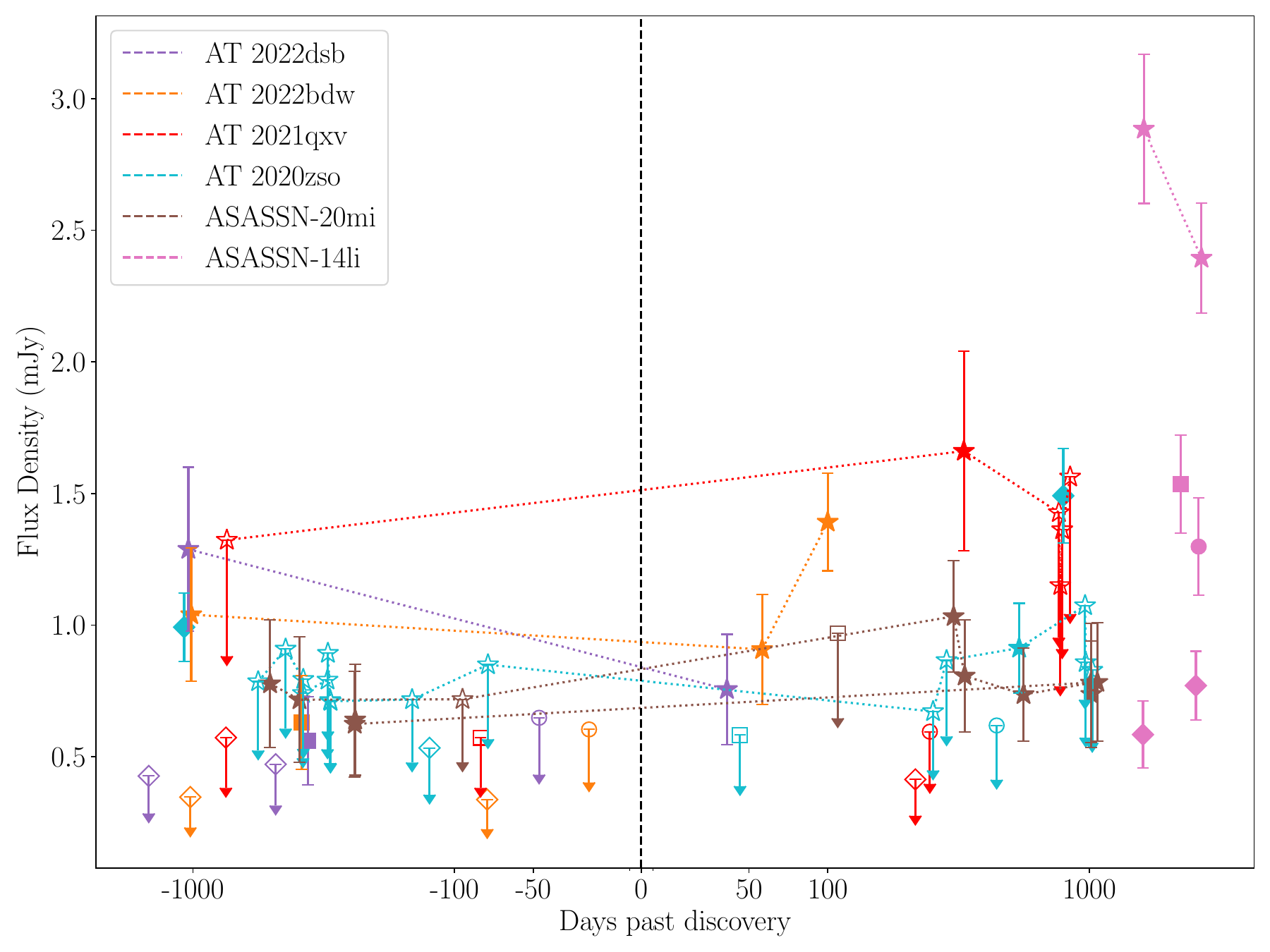}
    \caption{Lightcurves of the TDEs which did not show clear variability over the span of the RACS observations (roughly 3\,yrs) and might be contaminated by host emission: AT 2022dsb (purple), AT 2022bdw (orange), AT 2021qxv (red), AT 2020zso (cyan), ASASSN-20mi (brown) and ASASSN-14li (magenta). For each TDE the \rlow\ data are shown as stars, \rmid\ data as squares, \rhigh\ data as circles, and the VLASS data as diamonds. Filled markers represent detections and open markers represent upper limits. The black dashed line shows the optical outburst time. Where possible, we use RACS+VLASS measurements to estimate the spectral slope of the power law spectrum (see Section~\ref{sec:agns}).}
    \label{fig:tde-agns}
\end{figure*}

\begin{figure}
    \centering
    \includegraphics[scale=0.44]{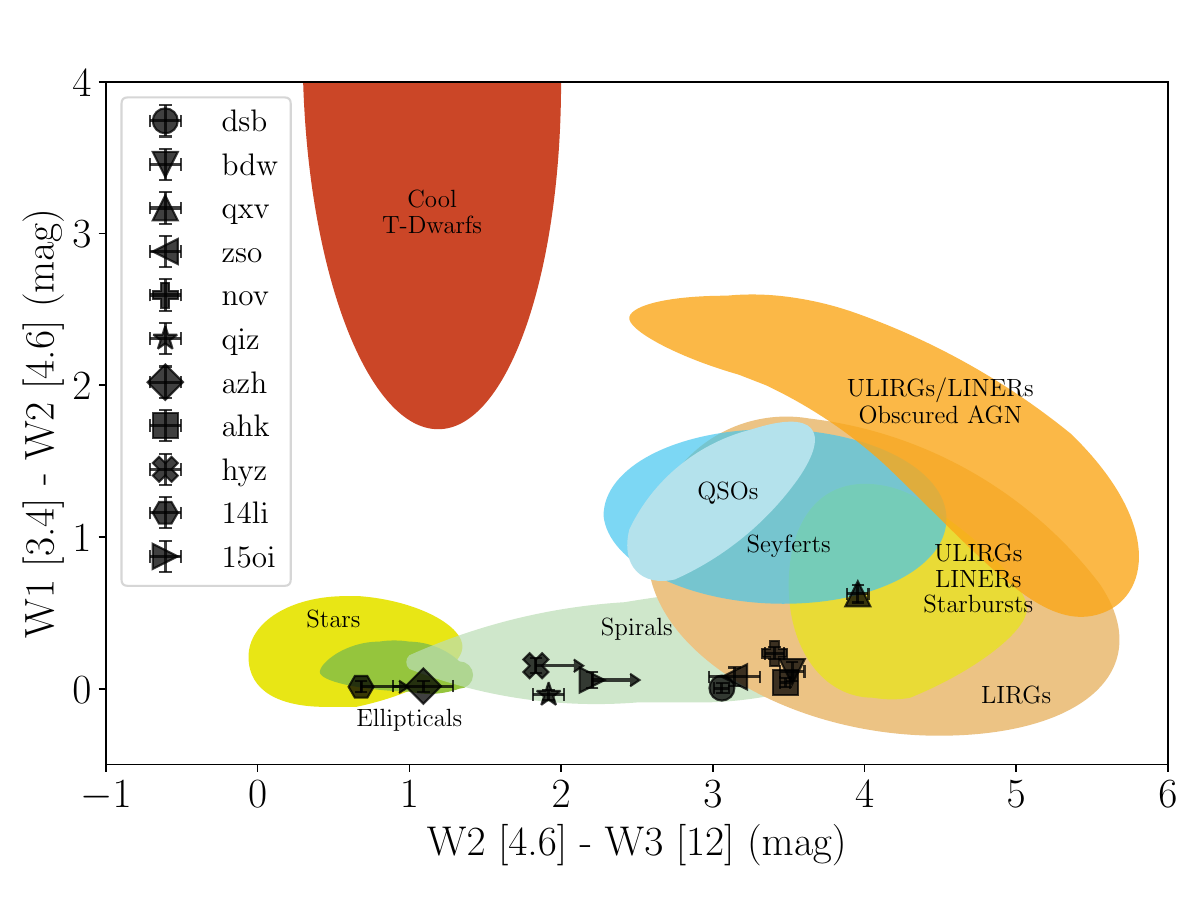}
    \caption{WISE color-color plot showing the magnitude difference in band 1 (W1) and band 2 (W2) on the Y axis, and band 2 (W2) and band 3 (W3) on the X axis. The contours for different object classes are adopted from \cite{wise}. We used a W1-W2 color of $>$ 0.8 to classify an object as AGN.}
    \label{fig:wise}
\end{figure}

In addition to the candidates where a rising/declining behavior is clearly seen, there are cases where the light curve showed little variation or was consistent with a non-varying source (the underlying host galaxy or AGN).  An AGN may be intrinsically variable, or variable due to external effects like scintillation \citep{Jauncey2016}. In both cases, if the flux density was consistent with a steady source within error bars between the two RACS epochs, we considered that to result from underlying AGN activity (clearly this is a conservative assumption, as we could be averaging over peaks or declines given our sparse sampling). Below we note such examples (see Figure~\ref{fig:tde-agns}). We cross-matched the TDEs in our sample with the WISE catalog \citep{wise} to look for AGN signature. Figure~\ref{fig:wise} shows the identified counterparts on a WISE  color-color plot \citep{wise}. We used a color difference of \textit{WISE band 1 (3.4\,$\mu$m) $-$ WISE band 2 (4.6\,$\mu$m)} $>$ 0.8 \citep{wise} to classify an object as an AGN. 

\begin{itemize}
    \item  \textbf{AT2020nov} was detected in both epochs of \rlow\ with no significant evolution between them, and also in \rmid\ and \rhigh\ (see Figure~\ref{fig:2020nov}).  The first \rlow\ observation pre-dated the optical outburst by $\sim400$\,days. We looked at the VLASS images and found that the same behavior was replicated at 3\,GHz. Recently, \cite{Cendes2023} also reported AT2020nov as probably dominated by an AGN in their study, with a non-evolving light curve at 6\,GHz. The lack of variability in the observed data seems to indicate that the radio emission is likely coming from the AGN activity itself. Exploiting the non-variability of this at different frequencies, we estimated the spectral index (see Figure~\ref{fig:2020nov}), assuming a power-law spectrum $S_{\nu} \propto \nu^{-\alpha}$ for AGN activity. Using the data from RACS, VLASS and \cite{Cendes2023} to find $\alpha=-0.64\pm0.04$, consistent with the typical AGN spectrum \citep{Condon2002,LoTUSSAGN}.
    \begin{figure}
        \centering
        \plotone{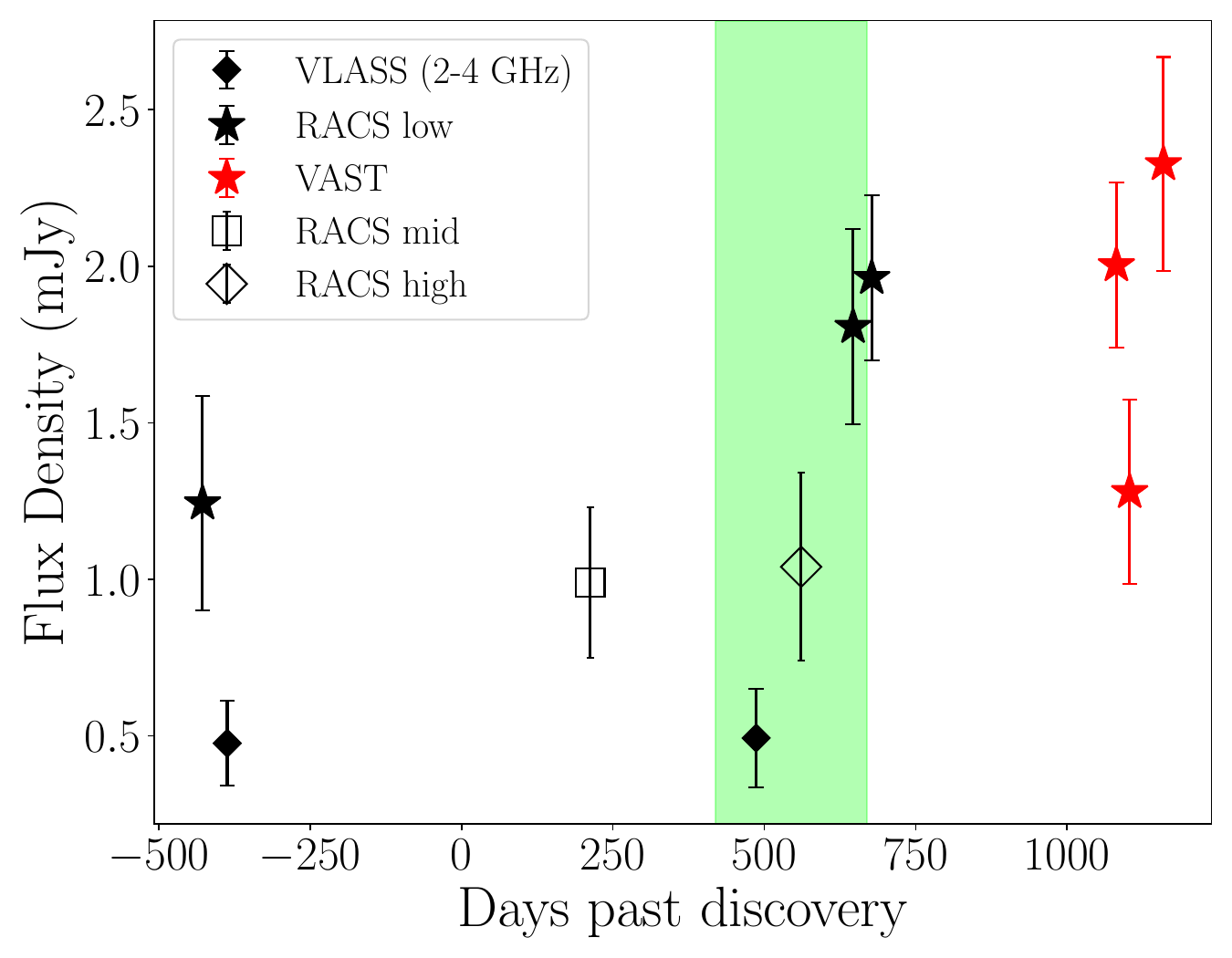}
        \caption{Light curve of TDE AT2020nov using RACS, VAST, and VLASS data, all resulting in detections, with very little variability. The green stripe shows the data points used to measure the spectral index of the power-law spectrum.}
    \label{fig:2020nov}
    \end{figure}
    \item \textbf{AT 2022dsb} was discovered by \cite{Stanek2022} on 2022~March~01 and had a radio detection reported by \cite{Goodwin2022dsb} roughly 20\,days later, but the transient nature of this source was not confirmed. \rlow\ epoch 1 had a \nsigma{4} pre-discovery measurement which points to an underlying AGN\footnote{There is a WISE counterpart within 1\arcsec\, of this position, but the WISE colors were not sufficiently conclusive to claim an AGN.} or host galaxy emission, a conclusion strengthened by the detection in epoch 2 which showed little variability roughly 60\,days after the optical outburst. There was also a pre-discovery measurement in \rmid. \rhigh\ and VLASS data resulted in upper limits (\nsigma{3}; 0.6\,mJy and 0.36\,mJy respectively). Based on these detections and upper limits, we estimated the spectral index to be $\alpha=-0.7\pm0.3$, typical of AGN.
    
    \item \textbf{AT 2022bdw} \citep{2022bdw}: No radio detection from this source has been reported so far, but we found pre-discovery detections in \rlow\ epoch 1 and \rmid\ data. Comparing it with \rlow\ epoch 2, which was post optical outburst, we found that the flux density level was consistent with a non-varying source, either the host AGN or host galaxy emission\footnote{This was one of the few fields that was observed twice as a part of \rlow\ epoch 2, separated by 45 days, and the flux density was consistent with a non-varying source to within \nsigma{2}.}. No emission was found in \rhigh\ or VLASS data and using these we estimate the spectral index of the background emission to be $\alpha=-0.8\pm0.3$, again typical of AGN.  

    \item \textbf{AT 2021qxv}: AT 2021qxv \citep{2021qxv} was observed as a part of the VAST survey in addition to the RACS survey. However, no strong detection has been found in any of the RACS/VAST data except for detection in \rlow\ epoch 2. \rmid\ showed a weak \nsigma{3} detection at this location, with \rhigh\,, VLASS and FIRST data resulting in null detections, and hence we conclude that the \rlow\ epoch 2 detection that we see is probably coming from an underlying AGN\footnote{WISE colors point to a probable AGN.} with a spectral index steeper than $\alpha=-1.1$.

    \item \textbf{AT 2020zso}: Discovered by \citet{Forster2020}, very weak radio detection of $22\pm7\,\mu Jy$ at 15\,GHz was reported roughly 1 month later by \cite{Alexander2021}, but following this, a null detection was made with the uGMRT \citep{2020zsoGMRT} at the central frequencies of 0.65 and 1.26\,GHz (upper limits of 46.6\,$\mu$Jy and 51.2\,$\mu$Jy). No strong detections were made in RACS/VAST data except for a single detection in \rlow\ epoch 2. VLASS data contains a \nsigma{5} detection in epoch 3.1, but the VAST observation that succeeded this resulted in non-detections so we cannot conclusively establish any late time transient activity from this source. It might be possible that the transient might take longer to rise at lower frequencies \citep{Chevalier1998} in which case future data from the VAST full survey will be very useful to check this. However, with the current data, we cannot rule out AGN variability.

    \item \textbf{ASASSN-14li}: ASASSN-14li \citep{ASASSN-14liatel,Holoien2016,Alexander2016} showed late-time fading that continued until $\sim 600$\,days in some bands \citep{Bright2018}. We found radio detections at this position in both \rlow\ epochs, but consistent with a steady flux density level. This source was also detected in \rmid\,, \rhigh\,, and VLASS, with no sign of evolution in the latter. Comparing the archival FIRST measurement $2.68\pm0.15$\,mJy at 1.4\,GHz with the \rmid\ observations indicate a $\sim 40$\% decrease in the flux density, indicating that the transient possibly faded away and we are looking at the variability from an AGN. Using RACS, and VLASS data, we find the spectral index $\alpha=-0.95\pm0.14$.
\end{itemize}

\section{Discussion}\label{sec:discussion}
\subsection{On the nature of detections}
\begin{figure}[!htb]
    \plotone{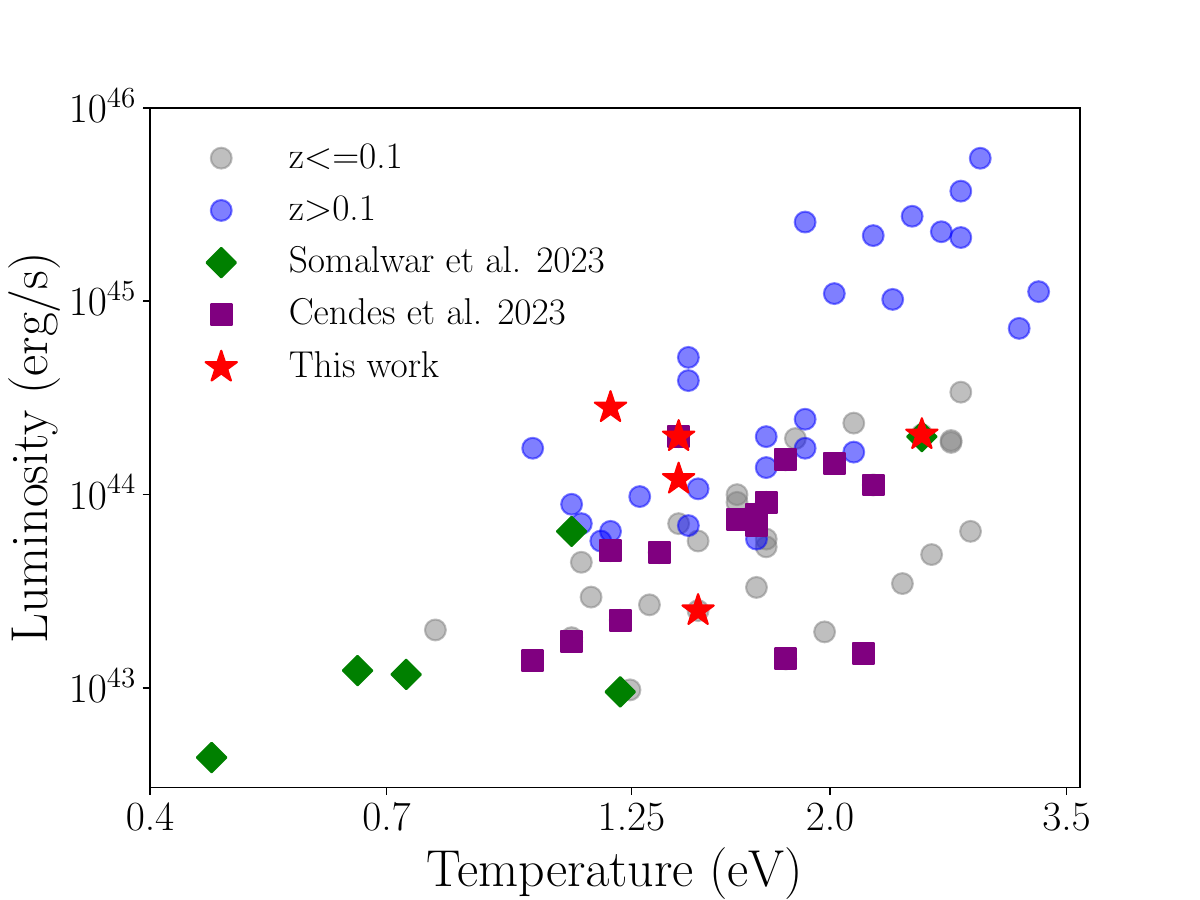}
    \caption{Blackbody luminosity vs temperature of the non-relativistic TDEs detected by various surveys. The background gray circles show the optical sample from the ZTF \citep{Hammerstein2023,Yao2023} that are within a redshift of z=0.1 and the blue circles show the TDE sample with z$>$0.1. The red stars show the optically-discovered TDEs that are detected in our radio sample. The purple squares show the optically discovered TDEs that resulted in radio detections in a targeted follow-up study by \citet{Cendes2023}. The green diamonds show the TDEs that are independently identified in radio by \citet{Somalwar2023}, but then confirmed optically.  Where multiple points overlap, sources are common to multiple samples.}
    \label{fig:source_dist}
\end{figure}

\begin{figure}
    \centering
    \plotone{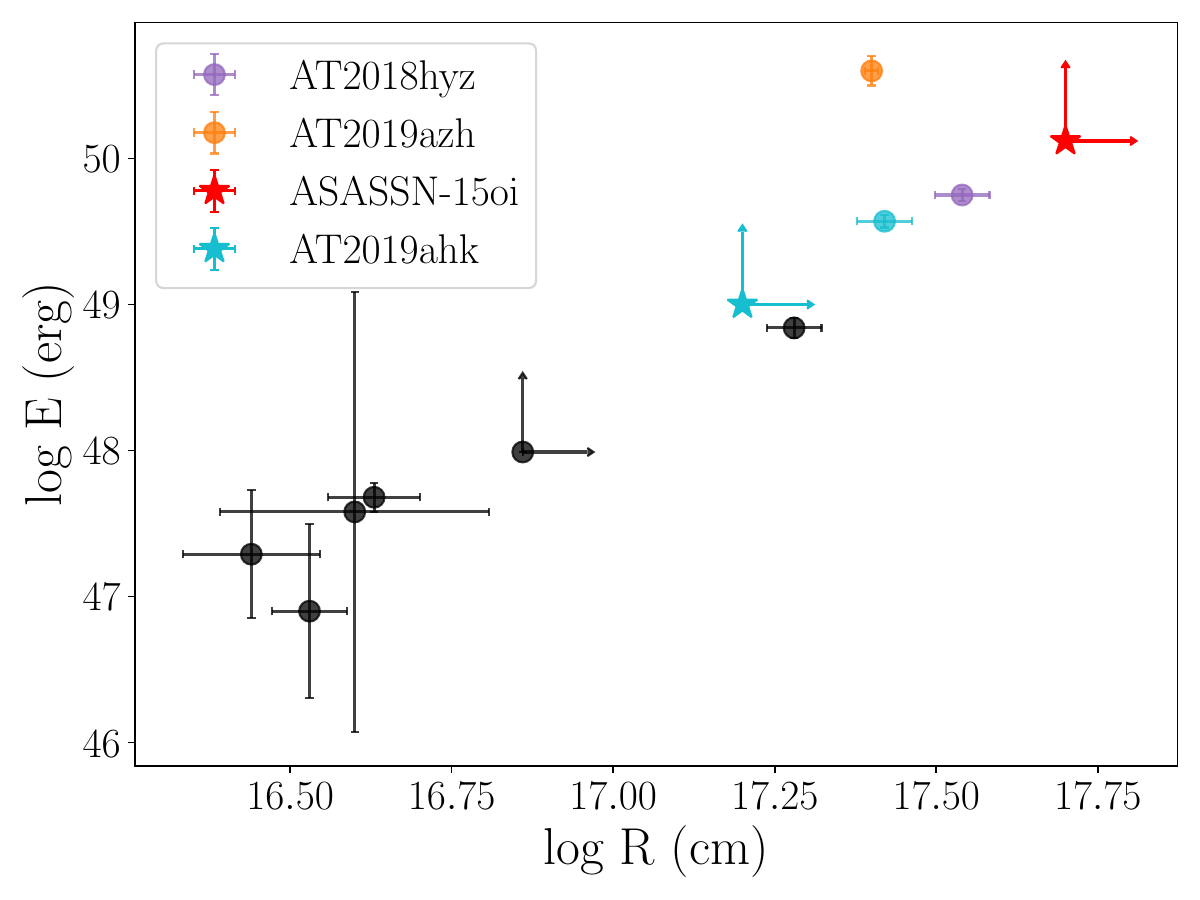}
    \caption{Estimates of equipartition radius and minimum total energy of the system, where possible, for the TDEs in our sample and archival TDEs (taken from \citealt{Cendes2023}). For our sample of TDEs, estimates derived from this study are shown as stars and those from the archival studies \citep{Horesh2021,Cendes2022,Christy2024} are shown as circles. Estimates for archival TDEs that show late-time activity are shown as black dots (adapted from \citealt{Cendes2023}).}
    \label{fig:equi-rad-en}
\end{figure}

Understanding the sample biases in all-sky searches is important in estimating the rates and expectations for future surveys. In particular, understanding if our radio-detected sample of TDEs forms an unbiased representation of the underlying optical population becomes important for future projections. Figure~\ref{fig:source_dist} shows the optical properties of the TDEs (blackbody luminosity vs temperature, as estimated from the optical data) that resulted in radio detections in the \rlow\ survey. Comparing the radio detections in optically discovered TDEs using the sample for this study and from \cite{Cendes2023}, we do not see preferential occupation of radio-detected TDEs in this phase space. We do see that there are no radio detections of TDEs with both high temperatures and luminosities (top right corner of the plot), but that can be attributed to the redshift because we do not expect detectable radio emission\footnote{This is under the expectation of detecting radio emission from a sub-relativistic outflow with $\nu L_{\nu}\approx 10^{38}$erg/s.} (given the current sensitivity limits of surveys like RACS/VLASS) from that sub-population. In the sub-sample of optical TDEs from which radio emission can be detected ($z \lesssim 0.1$), our sample, as well as the sample from \cite{Cendes2023}, is not biased towards certain classes of optical TDEs which suggests that the late-time detection of radio emission in TDEs might not be coming from a particular population of TDEs, but is a common feature of sub-relativistic TDEs in general. We then compare our estimates of the emission radius and minimum energy injected into the outflow with archival studies, under the equipartition situation (see Figure~\ref{fig:equi-rad-en}). We caution the reader that a strict comparison would need accurate modeling of the outflow expansion properties (linear/accelerating/decelerating) to compare estimates from different times. We hence restrict the sample to those that show late-time radio emission. Using the sample from \cite{Horesh2021,Cendes2022,Cendes2023} and \cite{Christy2024}, we find that our estimates for the emission radius and the energy injected are consistent with those reported in the literature.

\cite{Somalwar2023} did an untargeted search for TDEs using the first two epochs of VLASS data, and independently discovered radio-first detections of optically bright TDEs. These are shown as the green scatter in Figure~\ref{fig:source_dist}. While some of these seem to be consistent with the population of optically selected radio TDEs, \cite{Somalwar2023} suggests that some radio-discovered optically bright TDEs can have lower black body temperatures and luminosities which partly can be due to TDEs occuring in dusty environments. Data from the RACS survey, but also more importantly from the VAST survey, which has a cadence of $\sim 2$ months, should be very useful in conducting such untargeted searches. 

\subsection{Projections for the VAST survey}\label{sec:rates}
\begin{figure}
    \plotone{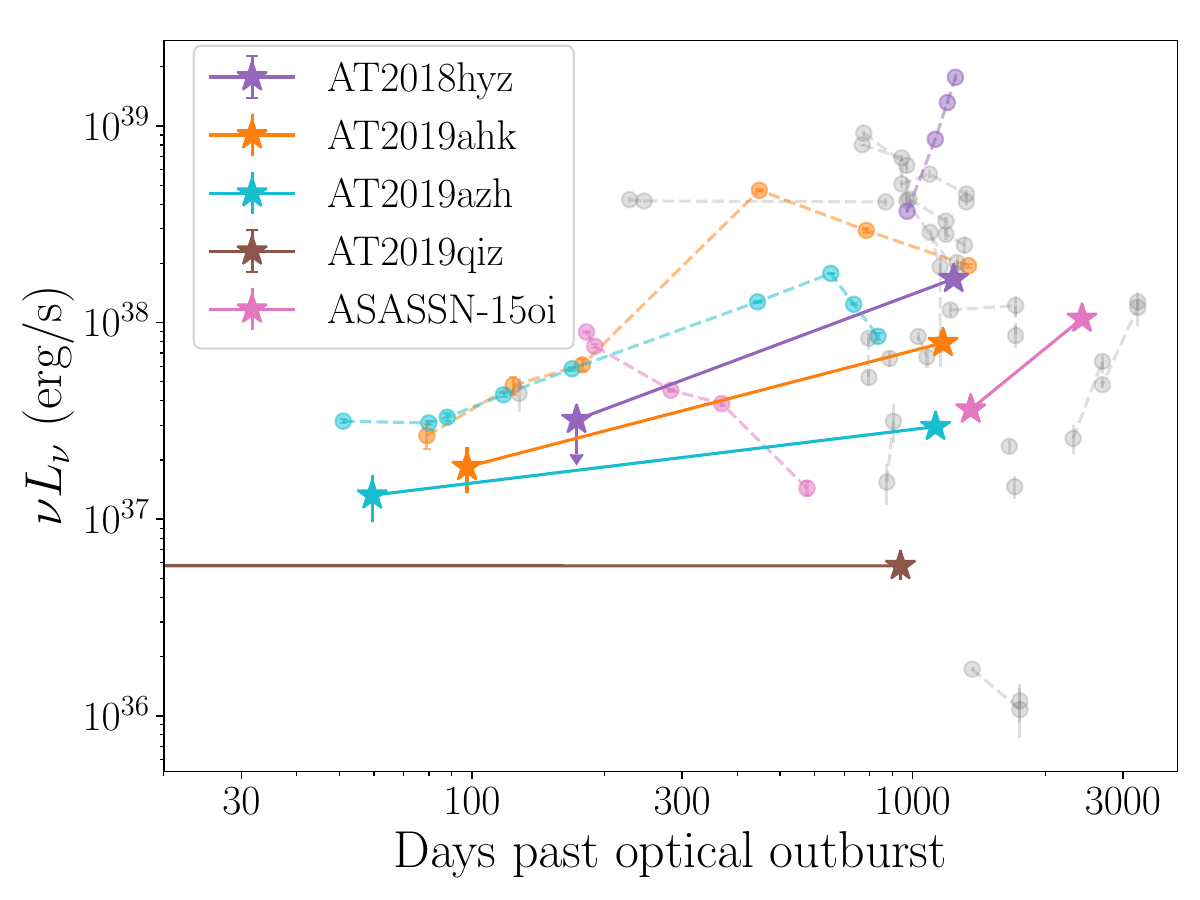}
    \caption{The radio luminosity of the 5 strong TDE detections in the RACS dataset at 887.5\,MHz. The solid line shows radio luminosity from RACS detections while the dashed lines show the from same 5-7\,GHz (data adapted from \citealt{Horesh2021,Cendes2022,Goodwin2022,Christy2024}). Shown in gray in the background are the light curves (at 6\,GHz) of the archival TDEs detected by \citet{Cendes2023} and \citet{katetdereview} from optically selected TDEs.}
    \label{fig:all_tde_lc}
\end{figure}

One of the important questions for an all-sky survey like RACS is the detection efficiency. Figure~\ref{fig:all_tde_lc} shows the radio luminosity at 887.5\,MHz of the TDEs detected in the RACS survey compared to those of the population. We see that all of these detections have $\nu L_{\nu} \approx 10^{38}$\,erg/s. We caution that comparison between light curves from our sample at 887.5\,MHz and archival light curves at 6\,GHz can be non-trivial and that the inferred radio luminosity $\nu L_{\nu}$ can have frequency dependence if $L_{\nu}$ does not exactly scale as $\nu^{-1}$. Thus, if the spectral index is steeper than $-1$, the radio luminosity, estimated from RACS will be an overestimation to the population (in Figure~\ref{fig:all_tde_lc}) and if shallower than $-1$, will be an underestimation. Despite this, if we assume $\nu L_{\nu} \approx 10^{38}$\,erg/s to be a typical estimate at 887.5\,MHz, then given the sensitivity of RACS survey (RMS noise of 0.25\,mJy/beam), then the survey should be complete out to $z=0.075$ ($d_L=350$\,Mpc). 

We then look at the total population of potentially detectable TDEs.  We require that they i) are in the RACS footprint ($<41\degree$ declination), ii) occurred before the \rlow\ epoch 2 (April 2022), and iii) are within $z=0.075$, which results in 23 TDEs.
Out of these, we detect 5 candidates where we are most likely seeing the afterglow and as many as 6 other events where we might be seeing the host AGN. Counting the 5 detections yields a (90\% confidence; estimated using \citealt{Gehrels1986}) detection rate of $22^{+15}_{-11}$\%. This is slightly more than but consistent within errors with \cite{katetdereview}, but slightly less than \cite{Cendes2023}, who finds late time radio activity in as much as 40\% of optically selected TDEs. It is worth mentioning here that, unlike a targeted search \citep[e.g.,][]{Cendes2023}, where continuous monitoring is done after initial detection, our results are based on observations roughly separated by 3\,yrs and hence we are completely insensitive to TDEs that rose and declined within this period, or to some of the most recent TDEs (that happened within a year of \rlow\ epoch 2), which are still rising, but are currently below our sensitivity threshold. Hence this detection efficiency can be considered as a conservative lower limit for future efforts: a survey with a longer duration and finer time sampling would be able to detect more sources at the same sensitivity threshold.

Recently \citet{Cendes2023} performed a comprehensive late-time follow-up of a sample of 23 TDEs and found radio emission lasting on timescales of $\sim$ a year in roughly 50\% of the TDEs. Using the first three years of optical data from ZTF, \citet{Yao2023} constrained the volumetric rate of optical TDEs to be $3.1^{+0.6}_{-1.0}\times 10^{-7}\, \rm Mpc^{-3}\, \rm yr^{-1}$. If we assume that as many as 50\% of optical TDEs (following \citealt{Cendes2023}) are capable of producing detectable late-time radio emission, then the current constraints on the rate of optical TDEs imply a rate of $1.5\times 10^{-7}\, \rm Mpc^{-3}\, \rm yr^{-1}$ for optically-selected, radio-emitting TDEs. This rate, coupled with the sensitivity of the VAST survey (RMS noise of 0.25\,mJy/beam), its footprint (roughly a quarter of the total sky), and the survey lifetime (4\,yrs), implies that VAST should be able to detect $\sim 20$ optically selected radio TDEs over the full survey. TDEs can also occur in highly dust-obscured environments, in which emission can be better studied at lower frequencies like the infrared where the emission can be powered by dust echoes \citep{Velzen2016,Ning2021,velzen2021} and at radio wavelengths, that need ambient material for the outflows/jets to interact with. The rate of the radio-bright optically-quiet TDEs is highly uncertain currently, particularly due to the lack of such studies. Hence the above-mentioned sample of $\sim 20$ TDEs can only be considered a lower limit on the detectable sample, given the current optical rate.

\subsection{RACS/VAST as a sub-GHz reference map}\label{sec:skymap}
Radio emission from TDEs can last $\sim$ years, and hence obtaining a robust host spectrum in the absence of one might imply that we need to wait for years before the transient fades away and the host galaxy dominates again. RACS, and in particular VAST, can be tremendously helpful in this respect since it provides a low-frequency (where the emission is brighter) reference image, that can be used to study the long-term variability (or lack thereof) of the host galaxy pre-explosion. To illustrate this, we provide the example of AT 2023clx where we can look for the pre-explosion radio emission using RACS data.

AT 2023clx was discovered by \cite{Stanek2023} on 2023~February~22, well after \rlow\ epoch 2. A radio detection was reported 4 days later by \cite{Sfaradi2023} consistent with the position of the optical transient. We found a persistent source in both epochs of \rlow\ data at the optical location and, using non-detections in VLASS, we constrain the radio spectrum of the host to be steeper than $-1.35$. 
For future observing campaigns that aim to do dedicated follow-up of these TDEs, RACS data can be very useful in estimating the level of host contamination. With the availability of VAST full survey data, not only radio first discoveries can be made, but also a well-sampled light curve with a cadence of 2 months leading up to the optical outburst, can be obtained.


\section{Conclusions}\label{sec:conclusions}
We conducted an untargeted search for radio emission in optically selected TDEs using data from the RACS survey, which resulted in 5 TDEs where the light curve showed significant evolution. For each of these TDEs, we modeled the evolution to show that the radio evolution at late times can undergo rebrightening and can be complex. We found that late-time activity can be quite common at radio wavelengths in sub-relativistic TDEs, adding to the sample presented by \cite{Cendes2023} who reached similar conclusions from targeted searches. Our search was based on the variability of the source over a timescale of roughly 3 years, which makes us insensitive to TDEs that evolve on timescales smaller than this, and we estimate the rate of optical TDEs in which late-time radio emission can be observed to be $22^{+15}_{-11}$\%. Using the current optical rates, we estimate a conservative lower limit on the number of TDEs that can be detected in the VAST survey to be $\sim 20$ over its survey span (4 years). 

\begin{acknowledgments}
We thank an anonymous referee for helpful comments.  AA and DLK are supported by NSF grant AST-1816492. Parts of this research were conducted by the Australian Research Council Centre of Excellence for Gravitational Wave Discovery (OzGrav), project number CE170100004. The Dunlap Institute is funded through an endowment established by the David Dunlap family and the University of Toronto. 
AH is grateful for the support by the the United States-Israel Binational Science Foundation (BSF grant 2020203) and by the  Sir Zelman Cowen Universities Fund. This research was supported by the Israel Science Foundation (grant No. 1679/23). 
HD acknowledges support from the Walter C. Sumner Memorial Fellowship and the Natural Sciences and Engineering Research Council of Canada (NSERC) through a Postgraduate Scholarship.
JP is supported by Australian Government Research Training Program Scholarships. KR thanks the LSST-DA Data Science Fellowship Program, which is funded by LSST-DA, the Brinson Foundation, and the Moore Foundation; Their participation in the program has benefited this work.
GRS is supported by NSERC Discovery Grant RGPIN-2021-0400.
This scientific work uses data obtained from Inyarrimanha Ilgari Bundara / the Murchison Radio-astronomy Observatory. 
The Australian SKA Pathfinder is part of the Australia Telescope National Facility (\url{https://ror.org/05qajvd42}) which is managed by CSIRO. Operation of ASKAP is funded by the Australian Government with support from the National Collaborative Research Infrastructure Strategy. ASKAP uses the resources of the Pawsey Supercomputing Centre. The establishment of ASKAP, the Murchison Radio-astronomy Observatory, and the Pawsey Supercomputing Centre are initiatives of the Australian Government, with support from the Government of Western Australia and the Science and Industry Endowment Fund.
\end{acknowledgments}

%
\newpage
\begin{longrotatetable}
\begin{deluxetable*}{lccllrcccc}
\tablecaption{Radio properties of the TDEs found in \rlow\ data set.\label{tab:flux}}
\tablehead{
\colhead{Name} & \colhead{RA} & \colhead{DEC} & \colhead{Discovery (UT)} & \colhead{z} & \colhead{$\delta$t} & \colhead{RACS low} & \colhead{RACS mid} & \colhead{RACS high} & \colhead{VLASS} \\
\colhead{} & \colhead{} & \colhead{} & \colhead{} & \colhead{} & \colhead{(days)} & \colhead{(mJy)} & \colhead{(mJy)} & \colhead{(mJy)} & \colhead{(mJy)}
}
\startdata
ASASSN-15oi & 2015-08-14 & 20h39m09.1s & $-$30d45m21s & 0.02 & 1355.1 & 4.21$\pm$0.27 & \nodata & \nodata & \nodata \\
 &  &  &  & \nodata & 1417.3 & \nodata & \nodata & \nodata & 8.95$\pm$1.35 \\
 &  &  &  & \nodata & 1961.3 & \nodata & 10.52$\pm$0.19 & \nodata & \nodata \\
 &  &  &  & \nodata & 2350.2 & \nodata & \nodata & 7.45$\pm$0.2 & \nodata \\
 &  &  &  & \nodata & 2377.7 & \nodata & \nodata & \nodata & 4.95$\pm$0.76 \\
 &  &  &  & \nodata & 2424.0 & 12.2$\pm$0.15 & \nodata & \nodata & \nodata \\
 &  &  &  & \nodata & 2860.9 & 7.28$\pm$0.17 & \nodata & \nodata & \nodata \\
 &  &  &  & \nodata & 2882.8 & 7.42$\pm$0.27 & \nodata & \nodata & \nodata \\
 &  &  &  & \nodata & 2938.6 & 6.51$\pm$0.2 & \nodata & \nodata & \nodata \\
AT2019ahk & 2019-01-29T21:50:24 & 07h00m11.5s & $-$66d02m24s & 0.026 & 97.5 & 1.24$\pm$0.33 & \nodata & \nodata & \nodata \\
 &  &  &  & \nodata & 738.6 & \nodata & 4.69$\pm$0.21 & \nodata & \nodata \\
 &  &  &  & \nodata & 1090.7 & \nodata & \nodata & 6.4$\pm$0.2 & \nodata \\
 &  &  &  & \nodata & 1172.5 & 5.87$\pm$0.29 & \nodata & \nodata & \nodata \\
AT2019azh & 2019-02-22T00:28:48 & 08h13m16.9s & 22d38m54s & 0.022 & 51.9 & \nodata & \nodata & \nodata & $<$0.58 \\
 &  &  &  & \nodata & 59.4 & 1.27$\pm$0.35 & \nodata & \nodata & \nodata \\
 &  &  &  & \nodata & 682.7 & \nodata & 3.88$\pm$0.19 & \nodata & \nodata \\
 &  &  &  & \nodata & 1025.3 & \nodata & \nodata & \nodata & 1.22$\pm$0.22 \\
 &  &  &  & \nodata & 1039.8 & \nodata & \nodata & 2.38$\pm$0.21 & \nodata \\
 &  &  &  & \nodata & 1127.5 & 2.83$\pm$0.2 & \nodata & \nodata & \nodata \\
AT2018hyz & 2018-11-06T15:21:36 & 10h06m50.9s & 01d41m34s & 0.046 & $-$304.2 & \nodata & \nodata & \nodata & $<$0.39 \\
 &  &  &  & \nodata & 172.9 & $<$0.69 & \nodata & \nodata & \nodata \\
 &  &  &  & \nodata & 294.4 & $<$0.78 & \nodata & \nodata & \nodata \\
 &  &  &  & \nodata & 357.2 & $<$0.81 & \nodata & \nodata & \nodata \\
 &  &  &  & \nodata & 358.2 & $<$0.75 & \nodata & \nodata & \nodata \\
 &  &  &  & \nodata & 408.2 & $<$0.73 & \nodata & \nodata & \nodata \\
 &  &  &  & \nodata & 430.1 & $<$0.75 & \nodata & \nodata & \nodata \\
 &  &  &  & \nodata & 431.1 & $<$0.85 & \nodata & \nodata & \nodata \\
 &  &  &  & \nodata & 436.1 & $<$0.82 & \nodata & \nodata & \nodata \\
 &  &  &  & \nodata & 437.1 & $<$0.68 & \nodata & \nodata & \nodata \\
 &  &  &  & \nodata & 438.1 & $<$0.61 & \nodata & \nodata & \nodata \\
 &  &  &  & \nodata & 591.7 & $<$0.67 & \nodata & \nodata & \nodata \\
 &  &  &  & \nodata & 661.5 & $<$0.94 & \nodata & \nodata & \nodata \\
 &  &  &  & \nodata & 676.2 & \nodata & \nodata & \nodata & 0.54$\pm$0.17 \\
 &  &  &  & \nodata & 795.2 & \nodata & 0.96$\pm$0.21 & \nodata & \nodata \\
 &  &  &  & \nodata & 988.6 & 1.23$\pm$0.2 & \nodata & \nodata & \nodata \\
 &  &  &  & \nodata & 1018.5 & 1.07$\pm$0.21 & \nodata & \nodata & \nodata \\
 &  &  &  & \nodata & 1158.2 & \nodata & \nodata & 4.85$\pm$0.22 & \nodata \\
 &  &  &  & \nodata & 1240.0 & 3.58$\pm$0.16 & \nodata & \nodata & \nodata \\
 &  &  &  & \nodata & 1553.8 & \nodata & \nodata & \nodata & 16.67$\pm$2.5 \\
 &  &  &  & \nodata & 1679.8 & 8.12$\pm$0.18 & \nodata & \nodata & \nodata \\
 &  &  &  & \nodata & 1701.7 & 8.43$\pm$0.19 & \nodata & \nodata & \nodata \\
 &  &  &  & \nodata & 1757.5 & 7.27$\pm$0.53 & \nodata & \nodata & \nodata \\
AT2019qiz & 2019-09-19T11:59:43 & 04h46m37.9s & $-$10d13m35s & 0.015 & $-$617.3 & \nodata & \nodata & \nodata & $<$0.36 \\
 &  &  &  & \nodata & $-$142.2 & $<$1.55 & \nodata & \nodata & \nodata \\
 &  &  &  & \nodata & 396.9 & \nodata & \nodata & \nodata & 1.1$\pm$0.26 \\
 &  &  &  & \nodata & 490.0 & \nodata & 1.33$\pm$0.17 & \nodata & \nodata \\
 &  &  &  & \nodata & 841.1 & \nodata & \nodata & 1.25$\pm$0.18 & \nodata \\
 &  &  &  & \nodata & 938.8 & 1.19$\pm$0.18 & \nodata & \nodata & \nodata \\
 &  &  &  & \nodata & 1282.5 & \nodata & \nodata & \nodata & 0.92$\pm$0.21 \\
 &  &  &  & \nodata & 1387.6 & $<$9.31 & \nodata & \nodata & \nodata \\
 &  &  &  & \nodata & 1445.4 & $<$2.11 & \nodata & \nodata & \nodata \\
AT2020zso & 2020-11-12T03:36:05.003 & 22h22m17.1s & $-$07d16m00s & 0.061 & $-$1080.0 & \nodata & \nodata & \nodata & 0.99$\pm$0.2 \\
 &  &  &  & \nodata & $-$563.2 & $<$0.79 & \nodata & \nodata & \nodata \\
 &  &  &  & \nodata & $-$442.4 & $<$0.91 & \nodata & \nodata & \nodata \\
 &  &  &  & \nodata & $-$379.6 & $<$0.74 & \nodata & \nodata & \nodata \\
 &  &  &  & \nodata & $-$378.7 & $<$0.79 & \nodata & \nodata & \nodata \\
 &  &  &  & \nodata & $-$305.9 & $<$0.79 & \nodata & \nodata & \nodata \\
 &  &  &  & \nodata & $-$304.9 & $<$0.89 & \nodata & \nodata & \nodata \\
 &  &  &  & \nodata & $-$298.8 & $<$0.71 & \nodata & \nodata & \nodata \\
 &  &  &  & \nodata & $-$297.8 & $<$0.71 & \nodata & \nodata & \nodata \\
 &  &  &  & \nodata & $-$145.3 & $<$0.72 & \nodata & \nodata & \nodata \\
 &  &  &  & \nodata & $-$124.6 & \nodata & \nodata & \nodata & $<$0.54 \\
 &  &  &  & \nodata & $-$74.5 & $<$0.85 & \nodata & \nodata & \nodata \\
 &  &  &  & \nodata & 46.2 & \nodata & $<$0.58 & \nodata & \nodata \\
 &  &  &  & \nodata & 252.7 & $<$0.67 & \nodata & \nodata & \nodata \\
 &  &  &  & \nodata & 284.5 & $<$0.87 & \nodata & \nodata & \nodata \\
 &  &  &  & \nodata & 442.1 & \nodata & \nodata & $<$0.62 & \nodata \\
 &  &  &  & \nodata & 538.8 & 0.91$\pm$0.17 & \nodata & \nodata & \nodata \\
 &  &  &  & \nodata & 794.8 & \nodata & \nodata & \nodata & 1.49$\pm$0.29 \\
 &  &  &  & \nodata & 962.7 & $<$1.07 & \nodata & \nodata & \nodata \\
 &  &  &  & \nodata & 967.7 & $<$0.86 & \nodata & \nodata & \nodata \\
 &  &  &  & \nodata & 1025.6 & $<$0.83 & \nodata & \nodata & \nodata \\
AT2021qxv & 2021-05-10T10:50:52.800 & 15h18m59.3s & $-$03d11m45s & 0.183 & $-$748.0 & \nodata & \nodata & \nodata & $<$0.59 \\
 &  &  &  & \nodata & $-$741.9 & $<$1.32 & \nodata & \nodata & \nodata \\
 &  &  &  & \nodata & $-$79.5 & \nodata & $<$0.57 & \nodata & \nodata \\
 &  &  &  & \nodata & 216.2 & \nodata & \nodata & \nodata & $<$0.41 \\
 &  &  &  & \nodata & 245.6 & \nodata & \nodata & $<$0.6 & \nodata \\
 &  &  &  & \nodata & 331.3 & 1.66$\pm$0.38 & \nodata & \nodata & \nodata \\
 &  &  &  & \nodata & 764.2 & $<$1.43 & \nodata & \nodata & \nodata \\
 &  &  &  & \nodata & 774.1 & $<$1.15 & \nodata & \nodata & \nodata \\
 &  &  &  & \nodata & 787.1 & $<$1.36 & \nodata & \nodata & \nodata \\
 &  &  &  & \nodata & 843.9 & $<$1.56 & \nodata & \nodata & \nodata \\
AT2022bdw & 2022-01-31T09:37:26.400 & 08h25m10.4s & 18d34m57s & 0.038 & $-$1023.4 & \nodata & \nodata & \nodata & $<$0.36 \\
 &  &  &  & \nodata & $-$1014.9 & 1.04$\pm$0.25 & \nodata & \nodata & \nodata \\
 &  &  &  & \nodata & $-$384.7 & \nodata & 0.63$\pm$0.18 & \nodata & \nodata \\
 &  &  &  & \nodata & $-$75.0 & \nodata & \nodata & \nodata & $<$0.35 \\
 &  &  &  & \nodata & $-$30.6 & \nodata & \nodata & $<$0.6 & \nodata \\
 &  &  &  & \nodata & 56.1 & 0.91$\pm$0.21 & \nodata & \nodata & \nodata \\
 &  &  &  & \nodata & 100.0 & 1.39$\pm$0.19 & \nodata & \nodata & \nodata \\
AT2022dsb & 2022-03-01T13:40:47 & 15h42m21.7s & $-$22d40m14s & 0.023 & $-$1475.1 & \nodata & \nodata & \nodata & $<$0.46 \\
 &  &  &  & \nodata & $-$1041.0 & 1.29$\pm$0.3 & \nodata & \nodata & \nodata \\
 &  &  &  & \nodata & $-$482.8 & \nodata & \nodata & \nodata & $<$0.48 \\
 &  &  &  & \nodata & $-$363.7 & \nodata & 0.56$\pm$0.16 & \nodata & \nodata \\
 &  &  &  & \nodata & $-$47.5 & \nodata & \nodata & $<$0.65 & \nodata \\
 &  &  &  & \nodata & 41.2 & 0.76$\pm$0.21 & \nodata & \nodata & \nodata \\
 ASASSN-14li & 2014-11-22T00:00:00 & 12h48m15.2s & 17d46m26s & 0.021 & 1602.4 & \nodata & \nodata & \nodata & 0.58$\pm$0.15 \\
 &  &  &  & \nodata & 1614.5 & 2.88$\pm$0.28 & \nodata & \nodata & \nodata \\
 &  &  &  & \nodata & 2230.0 & \nodata & 1.54$\pm$0.19 & \nodata & \nodata \\
 &  &  &  & \nodata & 2552.5 & \nodata & \nodata & \nodata & 0.77$\pm$0.17 \\
 &  &  &  & \nodata & 2615.9 & \nodata & \nodata & 1.3$\pm$0.18 & \nodata \\
 &  &  &  & \nodata & 2681.7 & 2.39$\pm$0.2 & \nodata & \nodata & \nodata \\
\enddata
\end{deluxetable*}
\end{longrotatetable}
\vspace{5mm}
\facilities{ASKAP}

\appendix
\startlongtable
\begin{deluxetable*}{l|r|cc|cc}\label{tab:limits}
\tablecaption{Upper limits (3-$\sigma$) on the radio emission from the RACS/VAST survey for the sample of TDEs that are in RACS footprint but resulted in non-detections.}
\tablehead{\colhead{Name} & \colhead{$\delta$t} & \multicolumn2c{Flux limit} & 
\multicolumn2c{Luminosity limit\tablenotemark{a}}\\
\hline
\colhead{} & \colhead{} & \colhead{RACS} & \colhead{VAST} & \colhead{RACS} & \colhead{VAST}\\
\hline
\colhead{} & \colhead{(days)} & \colhead{(mJy)} &\colhead{(mJy)} & \colhead{(ergs/s)} & \colhead{(ergs/s)}}
\startdata
AT2016fnl & 1121 & 0.8 & \nodata & 4.6$\times 10^{36}$ & \nodata \\
AT2016fnl & 2047 & 0.6 & \nodata & 3.5$\times 10^{36}$ & \nodata \\
AT2018dyb & 466 & 2.1 & \nodata & 1.5$\times 10^{37}$ & \nodata \\
AT2018dyb & 1370 & 2.4 & \nodata & 1.6$\times 10^{37}$ & \nodata \\
AT2018fyk & 419 & 0.6 & \nodata & 5.0$\times 10^{37}$ & \nodata \\
AT2018fyk & 517 & \nodata & 0.7 & \nodata & 5.4$\times 10^{37}$ \\
AT2018fyk & 518 & \nodata & 0.5 & \nodata & 4.4$\times 10^{37}$ \\
AT2018fyk & 520 & \nodata & 0.5 & \nodata & 4.2$\times 10^{37}$ \\
AT2018fyk & 521 & \nodata & 0.5 & \nodata & 4.3$\times 10^{37}$ \\
AT2018fyk & 727 & \nodata & 0.6 & \nodata & 4.9$\times 10^{37}$ \\
AT2018fyk & 1024 & \nodata & 0.5 & \nodata & 3.8$\times 10^{37}$ \\
AT2018fyk & 1049 & \nodata & 0.4 & \nodata & 3.7$\times 10^{37}$ \\
AT2018fyk & 1080 & \nodata & 0.5 & \nodata & 3.7$\times 10^{37}$ \\
AT2018fyk & 1334 & 0.4 & \nodata & 3.3$\times 10^{37}$ & \nodata \\
AT2018fyk & 1762 & \nodata & 0.7 & \nodata & 5.4$\times 10^{37}$ \\
AT2018fyk & 1763 & \nodata & 0.5 & \nodata & 3.9$\times 10^{37}$ \\
AT2018hco & 355 & 1.1 & \nodata & 2.2$\times 10^{38}$ & \nodata \\
AT2018hco & 1282 & 0.9 & \nodata & 1.6$\times 10^{38}$ & \nodata \\
AT2018iih & 320 & 0.7 & \nodata & 9.1$\times 10^{38}$ & \nodata \\
AT2018iih & 1241 & 0.5 & \nodata & 6.2$\times 10^{38}$ & \nodata \\
AT2018lna & 271 & 0.8 & \nodata & 1.6$\times 10^{38}$ & \nodata \\
AT2018lna & 1193 & 0.6 & \nodata & 1.2$\times 10^{38}$ & \nodata \\
AT2018zr & 571 & 0.7 & \nodata & 7.8$\times 10^{37}$ & \nodata \\
AT2018zr & 1491 & 0.6 & \nodata & 6.5$\times 10^{37}$ & \nodata \\
AT2019bhf & 245 & 1.7 & \nodata & 6.1$\times 10^{38}$ & \nodata \\
AT2019bhf & 1147 & 0.8 & \nodata & 2.9$\times 10^{38}$ & \nodata \\
AT2019dsg & 191 & 0.9 & \nodata & 5.6$\times 10^{37}$ & \nodata \\
AT2019dsg & 1095 & 0.6 & \nodata & 3.6$\times 10^{37}$ & \nodata \\
AT2019gte & 139 & 0.8 & \nodata & 1.5$\times 10^{38}$ & \nodata \\
AT2019gte & 252 & \nodata & 0.8 & \nodata & 1.5$\times 10^{38}$ \\
AT2019gte & 252 & \nodata & 1.0 & \nodata & 1.7$\times 10^{38}$ \\
AT2019gte & 254 & \nodata & 1.0 & \nodata & 1.8$\times 10^{38}$ \\
AT2019gte & 255 & \nodata & 0.7 & \nodata & 1.2$\times 10^{38}$ \\
AT2019gte & 255 & \nodata & 0.9 & \nodata & 1.5$\times 10^{38}$ \\
AT2019gte & 462 & \nodata & 0.8 & \nodata & 1.4$\times 10^{38}$ \\
AT2019gte & 760 & \nodata & 0.6 & \nodata & 1.1$\times 10^{38}$ \\
AT2019gte & 783 & \nodata & 0.7 & \nodata & 1.1$\times 10^{38}$ \\
AT2019gte & 814 & \nodata & 0.7 & \nodata & 1.2$\times 10^{38}$ \\
AT2019gte & 1037 & 0.6 & \nodata & 1.0$\times 10^{38}$ & \nodata \\
AT2019gte & 1484 & \nodata & 0.6 & \nodata & 1.1$\times 10^{38}$ \\
AT2019gte & 1497 & \nodata & 0.9 & \nodata & 1.6$\times 10^{38}$ \\
AT2019lwu & 86 & 0.8 & \nodata & 2.8$\times 10^{38}$ & \nodata \\
AT2019lwu & 197 & \nodata & 0.9 & \nodata & 3.0$\times 10^{38}$ \\
AT2019lwu & 198 & \nodata & 0.9 & \nodata & 2.9$\times 10^{38}$ \\
AT2019lwu & 200 & \nodata & 0.8 & \nodata & 2.8$\times 10^{38}$ \\
AT2019lwu & 200 & \nodata & 0.9 & \nodata & 3.0$\times 10^{38}$ \\
AT2019lwu & 201 & \nodata & 0.7 & \nodata & 2.3$\times 10^{38}$ \\
AT2019lwu & 201 & \nodata & 0.7 & \nodata & 2.5$\times 10^{38}$ \\
AT2019lwu & 407 & \nodata & 0.9 & \nodata & 3.0$\times 10^{38}$ \\
AT2019lwu & 706 & \nodata & 0.8 & \nodata & 2.7$\times 10^{38}$ \\
AT2019lwu & 730 & \nodata & 0.8 & \nodata & 2.7$\times 10^{38}$ \\
AT2019lwu & 760 & \nodata & 0.7 & \nodata & 2.5$\times 10^{38}$ \\
AT2019lwu & 1001 & 0.6 & \nodata & 1.9$\times 10^{38}$ & \nodata \\
AT2019lwu & 1444 & \nodata & 0.7 & \nodata & 2.5$\times 10^{38}$ \\
AT2019lwu & 1444 & \nodata & 0.8 & \nodata & 2.7$\times 10^{38}$ \\
AT2019vcb & $-$52 & 0.8 & \nodata & 1.4$\times 10^{38}$ & \nodata \\
AT2019vcb & 868 & 0.5 & \nodata & 1.0$\times 10^{38}$ & \nodata \\
AT2020acka & $-$421 & 3.3 & \nodata & 1.2$\times 10^{40}$ & \nodata \\
AT2020acka & 479 & 1.3 & \nodata & 4.5$\times 10^{39}$ & \nodata \\
AT2020neh & $-$247 & 0.9 & \nodata & 7.8$\times 10^{37}$ & \nodata \\
AT2020neh & 654 & 0.6 & \nodata & 5.3$\times 10^{37}$ & \nodata \\
AT2020pj & $-$79 & 0.6 & \nodata & 6.4$\times 10^{37}$ & \nodata \\
AT2020pj & 822 & 0.5 & \nodata & 5.2$\times 10^{37}$ & \nodata \\
AT2020vwl & $-$361 & 0.7 & \nodata & 2.0$\times 10^{37}$ & \nodata \\
AT2020vwl & 540 & 0.5 & \nodata & 1.4$\times 10^{37}$ & \nodata \\
AT2021ack & $-$455 & 1.1 & \nodata & 4.7$\times 10^{38}$ & \nodata \\
AT2021ack & 447 & 1.0 & \nodata & 4.5$\times 10^{38}$ & \nodata \\
AT2021ack & 484 & 0.8 & \nodata & 3.7$\times 10^{38}$ & \nodata \\
AT2021ack & 885 & \nodata & 1.1 & \nodata & 5.1$\times 10^{38}$ \\
AT2021ack & 905 & \nodata & 1.3 & \nodata & 5.8$\times 10^{38}$ \\
AT2021axu & $-$481 & 0.6 & \nodata & 6.2$\times 10^{38}$ & \nodata \\
AT2021axu & 439 & 0.5 & \nodata & 5.0$\times 10^{38}$ & \nodata \\
AT2021blz & $-$472 & 0.7 & \nodata & 3.3$\times 10^{37}$ & \nodata \\
AT2021blz & 445 & 0.5 & \nodata & 2.4$\times 10^{37}$ & \nodata \\
AT2021blz & 886 & \nodata & 0.5 & \nodata & 2.3$\times 10^{37}$ \\
AT2021ehb & $-$487 & 0.8 & \nodata & 5.3$\times 10^{36}$ & \nodata \\
AT2021ehb & 397 & 0.7 & \nodata & 4.9$\times 10^{36}$ & \nodata \\
AT2021gje & $-$504 & 1.0 & \nodata & 4.3$\times 10^{39}$ & \nodata \\
AT2021gje & 380 & 0.6 & \nodata & 2.5$\times 10^{39}$ & \nodata \\
AT2021jjm & $-$540 & 1.3 & \nodata & 7.8$\times 10^{38}$ & \nodata \\
AT2021jjm & 365 & 0.7 & \nodata & 4.1$\times 10^{38}$ & \nodata \\
AT2021jsg & $-$471 & 0.6 & \nodata & 2.3$\times 10^{38}$ & \nodata \\
AT2021jsg & 449 & 0.5 & \nodata & 2.0$\times 10^{38}$ & \nodata \\
AT2021lo & $-$446 & 1.1 & \nodata & 6.4$\times 10^{38}$ & \nodata \\
AT2021lo & 460 & 0.7 & \nodata & 3.9$\times 10^{38}$ & \nodata \\
AT2021lo & 894 & \nodata & 0.7 & \nodata & 3.9$\times 10^{38}$ \\
AT2021lo & 913 & \nodata & 0.7 & \nodata & 4.3$\times 10^{38}$ \\
AT2021mhg & $-$595 & 0.8 & \nodata & 9.4$\times 10^{37}$ & \nodata \\
AT2021mhg & 331 & 0.6 & \nodata & 7.4$\times 10^{37}$ & \nodata \\
AT2021uqv & $-$674 & 0.8 & \nodata & 2.2$\times 10^{38}$ & \nodata \\
AT2021uqv & 253 & 0.9 & \nodata & 2.4$\times 10^{38}$ & \nodata \\
AT2021uvz & $-$656 & 0.8 & \nodata & 6.6$\times 10^{38}$ & \nodata \\
AT2021uvz & 265 & 0.6 & \nodata & 5.2$\times 10^{38}$ & \nodata \\
AT2021yte & $-$719 & 0.8 & \nodata & 5.1$\times 10^{37}$ & \nodata \\
AT2021yte & 218 & 0.6 & \nodata & 3.7$\times 10^{37}$ & \nodata \\
AT2021yte & 234 & 0.5 & \nodata & 3.4$\times 10^{37}$ & \nodata \\
AT2021yzv & $-$655 & 0.9 & \nodata & 2.3$\times 10^{39}$ & \nodata \\
AT2021yzv & 228 & 0.8 & \nodata & 2.0$\times 10^{39}$ & \nodata \\
AT2022adm & $-$829 & 1.6 & \nodata & 1.4$\times 10^{38}$ & \nodata \\
AT2022adm & 72 & 0.7 & \nodata & 6.3$\times 10^{37}$ & \nodata \\
AT2022arb & $-$833 & 1.0 & \nodata & 8.7$\times 10^{37}$ & \nodata \\
AT2022arb & $-$725 & \nodata & 1.0 & \nodata & 8.8$\times 10^{37}$ \\
AT2022arb & $-$721 & \nodata & 1.1 & \nodata & 9.6$\times 10^{37}$ \\
AT2022arb & $-$721 & \nodata & 1.2 & \nodata & 1.0$\times 10^{38}$ \\
AT2022arb & $-$720 & \nodata & 0.9 & \nodata & 8.1$\times 10^{37}$ \\
AT2022arb & $-$720 & \nodata & 1.3 & \nodata & 1.2$\times 10^{38}$ \\
AT2022arb & $-$719 & \nodata & 4.0 & \nodata & 3.5$\times 10^{38}$ \\
AT2022arb & $-$718 & \nodata & 0.8 & \nodata & 7.4$\times 10^{37}$ \\
AT2022arb & $-$718 & \nodata & 1.2 & \nodata & 1.1$\times 10^{38}$ \\
AT2022arb & $-$703 & \nodata & 3.9 & \nodata & 3.4$\times 10^{38}$ \\
AT2022arb & $-$512 & \nodata & 3.8 & \nodata & 3.4$\times 10^{38}$ \\
AT2022arb & $-$213 & \nodata & 0.9 & \nodata & 7.9$\times 10^{37}$ \\
AT2022arb & $-$190 & \nodata & 0.7 & \nodata & 6.5$\times 10^{37}$ \\
AT2022arb & $-$160 & \nodata & 0.9 & \nodata & 7.5$\times 10^{37}$ \\
AT2022arb & 66 & 0.6 & \nodata & 5.4$\times 10^{37}$ & \nodata \\
AT2022arb & 504 & \nodata & 0.7 & \nodata & 6.0$\times 10^{37}$ \\
AT2022arb & 524 & \nodata & 0.8 & \nodata & 7.5$\times 10^{37}$ \\
AT2022czy & $-$843 & 0.9 & \nodata & 2.7$\times 10^{38}$ & \nodata \\
AT2022czy & 39 & 0.7 & \nodata & 1.9$\times 10^{38}$ & \nodata \\
AT2022dyt & $-$885 & 0.7 & \nodata & 8.3$\times 10^{37}$ & \nodata \\
AT2022dyt & 33 & 0.6 & \nodata & 7.0$\times 10^{37}$ & \nodata \\
AT2022exr & $-$887 & 0.7 & \nodata & 1.6$\times 10^{38}$ & \nodata \\
AT2022exr & 14 & 0.6 & \nodata & 1.2$\times 10^{38}$ & \nodata \\
\enddata
\tablenotetext{a}{Radio luminosity is estimated as $\nu L_{\nu}$ and we assume the emission to be spherical which results in the inclusion of 4$\pi$ factor.}
\end{deluxetable*}


          
\bibliography{references}{}
\bibliographystyle{aasjournal}

\end{document}